\documentclass[12pt]{iopart}
\usepackage{amssymb}
\usepackage{braket}
\usepackage[backend=biber, style=numeric, sorting=none, maxcitenames=50, maxbibnames=99, minbibnames=10]{biblatex}
\usepackage{graphicx}
\usepackage{booktabs}
\usepackage{caption}
\usepackage{simpler-wick}
\usepackage{multirow}

\expandafter\let\csname equation*\endcsname\relax
\expandafter\let\csname endequation*\endcsname\relax
\usepackage{amsmath,amssymb}

\begin{document}

\title[Reconstructing Conformal Field Theoretical Compositions]{Reconstructing Conformal Field Theoretical Composition with Transformers}

\author{Haotian Cao$^{a,b}$, Garrett W. Merz$^{a,b}$, Kyle Cranmer$^{a,b}$, Gary Shiu$^{a}$}

\address{$^{a}$ Department of Physics, University of Wisconsin-Madison}
\address{$^{b}$ Data Science Institute, University of Wisconsin-Madison}
\ead{hcao39@wisc.edu, garrett.merz@wisc.edu, kyle.cranmer@wisc.edu, gshiu@wisc.edu}
\vspace{10pt}
\begin{indented}
\item[]May 2026
\end{indented}

\begin{abstract}
We study the use of transformers to reconstruct the compositions of tensor products of two-dimensional rational conformal field theories (RCFTs) based on their low-energy spectra. The task is challenging due to its combinatorial nature. 
The constituent theories are characterized by their central charges and affine Lie algebra labels. We achieve $98\%$ accuracy in recovering the constituents of tensor products theories constructed from Wess-Zumino-Witten models. We further demonstrate that our method generalizes to CFTs with larger central charge and unseen classes of RCFTs by adding a small number of out-of-domain examples. Our results show that transformers are effective at this task and point towards a new tool for 
bulk reconstruction in AdS/CFT. 
\end{abstract}

%
%
%
%
%

\section{Introduction}


Conformal Field Theories (CFTs) are scale-invariant Quantum Field Theories (QFTs) with conformal symmetry, which places strong constraints on their dynamics, especially in two dimensions \cite{orange_book,applied_cft}. They appear across physics: in statistical mechanics, CFTs describe scale-invariant systems such as spin lattices at criticality \cite{patashinskii1979fluctuation}, or, in string theory, the worldsheet dynamics of strings is governed by a two-dimensional CFT.

CFTs are also of theoretical interest due to the AdS/CFT correspondence \cite{Maldacena:1997re} that connects a $d+1$-dimensional gravitational theory to a $d$-dimensional CFT on the boundary without gravity. 
In the context of AdS/CFT, the bulk geometric data are encoded indirectly in the boundary CFT observables, with low-lying operator spectra playing a prominent role. Therefore, if one can reliably infer the constituent structure of a CFT such as its central charges\footnote{The central charge, $c$, is a key quantity of a CFT that measures the effective number of degrees of freedom.} and symmetry algebra based on the low-energy behavior, we will have sharper tools for reconstruction of bulk gravitational spacetimes from boundary information \cite{Hashimoto2024, dl_and_ads/qcd}. 

This motivates the studies presented in this paper. We explore if deep learning techniques can recover the central charges and symmetry algebra of a CFT based on the low-energy behavior. As a starting point, we consider Rational Conformal Field Theories (RCFTs), which are a particularly tractable subclass of CFTs that possess a finite number of so-called \textit{primary operators}.
These RCFTs are widely used in exactly solvable constructions of 
string compactifications \cite{Polchinski_vol2}. 
Furthermore, in the context of AdS/CFT,
the low-lying spectra of RCFTs 
are 
a key ingredient for finding their gravity duals \cite{gravity_dual_ising, average_narain_moduli, mukhi_average_rcft}. 
RCFTs with 
large central charge (one of the necessary conditions for finding Einstein gravity duals) 
can be constructed by forming tensor products of well-studied RCFTs with small central charge, including minimal models characterized by the Virasoro algebra \cite{BELAVIN1984333} and Wess-Zumino-Witten (WZW) models characterized by affine Kac-Moody algebras \cite{orange_book}. 


In addition, the principle of tensor product models underlies Gepner models\footnote{An example of Gepner models is a tensor product of $5$ $\mathcal{N}=2$ superconformal minimal models \cite{Gepner_spacetime_susy} with total central charge $9$.
This Gepner model describes a six-dimensional compactified geometry known as a Calabi-Yau manifold at a singular point of the moduli space.} \cite{Gepner_spacetime_susy,Gepner_exactly_solvable}, which yield exactly solvable worldsheet constructions of Calabi–Yau compactifications (see \cite{Marchesano:2024gul} for a recent review).  Moreover, RCFTs provide an infinite family of solvable models. For a bounded central charge, there are finitely many of them which can be classified \cite{Mukhi_2023,Sunil_Sen_1988}.
Understanding how to map from the low-lying spectra to the underlying constituent CFTs may help us uncover 
mathematical structures of this infinite theory landscape.

We frame our task as sequence-to-sequence task and use a transformer to solve an inverse problem. Our {\bf forward process} involves forming a tensor product from a set of constituent CFTs and then calculating its low-lying spectra. The forward process is straightforward because central charges and operator dimensions are additive -- though there is a combinatorial element that makes the inverse problem difficult. The {\bf inverse problem} is to recover the constituent CFTs from the low-lying spectra, which is much harder, as distinct tensor products can produce nearly identical low-energy behavior.  Here, the input low-lying energy spectra are the primary conformal dimensions smaller than $1$, and we characterize the output decomposition as the central charges and symmetry algebra labels of each constituent CFT. Because the conformal dimensions and central charges are rational numbers (ratios of integers), we can tokenize the input as the ratios of integers. The naive, brute-force approach to this inverse problem scales like $|\mathcal{C}|^L$, where $|\mathcal{C}|$ denotes the cardinality of the set of constituent CFTs $\mathcal{C}$ and $L$ denotes the number of constituent CFTs in the tensor product. This exponential scaling motivates a more efficient and scalable approach based on sequence-to-sequence modeling.

In this paper, we start by reviewing the necessary ingredients of CFTs to generate our dataset in section \ref{sec: basics}. Then, in Section \ref{sec: data_generation}, we discuss the generation of our training data.
We also discuss our implementation details for tokenization and Transformer model in Section \ref{sec: implement}. In Section \ref{sec: base_model_training}, we study the performance of the Transformers for the inverse problem. In Section \ref{sec: c-gen} and Section \ref{sec: unseen_class}, we demonstrate the capability of Transformers to generalize to different datasets. In \ref{sec: sugawara}, \ref{sec: appendix_spectra}, and \ref{sec: coset+tensor}, we review the detailed derivations that are necessary to generate our datasets. In \ref{sec: learning_dynamics}, we present the learning dynamics of Transformers.

\subsection{Related Work}
In recent years, Machine Learning (ML) and Artificial Intelligence (AI) have accelerated the study of many scientific problems. Among various architectures, Transformers that employ an attention mechanism \cite{attentionneed} are promising and powerful tools that have revolutionized Natural Language Processing (NLP) \cite{bert} and computer vision \cite{cv}. Transformers have influenced fields in the basic sciences. In mathematics, transformers and Large Language Models have boosted mathematical research in various domains as seen in \cite{alfarano2024globallyapunovfunctionslongstanding}, \cite{charton2024patternboostconstructionsmathematicslittle}, and \cite{funsearch}. Moreover, Transformers have helped different research areas of theoretical high energy physics \cite{Cai:2024znx, Cheung:2024svk, new_calabi-yau2025, holography-transformer, simplify-polylog}.



There have been several studies that apply ML methods to understand the low-lying spectra of CFTs. Work using Principal Component Analysis (PCA) has shown that for a $3D$ $\mathcal{N}=2$ strongly coupled supersymmetric theory, 
the lowest energy gaps capture most of the information in the energy spectra \cite{Gukov2024bps}. Moreover, Kuo et al \cite{kuo2021decodingcft} showed that one can use ML to 
classify $2D$ RCFTs and identify the 
nature and the value of critical points of several strongly coupled spin systems based on the lowest $20$ energy levels. In addition, \cite{ml_etudes_cft} and \cite{solving_cft_with_ai} both applied ML techniques such as deep neural networks and Reinforcement Learning (RL) to study other properties of $2D$ or $3D$ CFTs such as correlation functions and OPE coefficients. The investigation of energy spectra of $2D$ RCFTs in symbolic forms 
was
performed in \cite{cao2024learningCFTpysr} with a symbolic regression package in python called pySR \cite{cranmer2023pysr}. More recently, Benjamin et al \cite{benjamin2026descendingmodularbootstrap} investigated the landscape of two-dimensional CFTs by searching for numerical solutions to the modular bootstrap equation with stochastic sampling optimization.

\section{The Basics of Conformal Field Theory} \label{sec: basics}

Similar to statistical physics, the spectra of CFTs are embedded in the partition functions as conformal ``tower" of states. The modular invariant partition function that contains holomorphic (left-moving) and anti-holomorphic (right-moving) sectors takes the general sesquilinear form 
\begin{equation} \label{eqn: partition_function}
    Z(\tau) = \sum_{j\bar{j}}N_{j\bar{j}}\chi_j (q) \bar{\chi}_{\bar{j}}(\bar{q})
\end{equation}
where $q = e^{2\pi i\tau}$ and $\chi_j(q) = \mathrm{Tr} \ q^{L_0-\frac{c}{24}}$ is the generating function with central charge $c$. We see that the energy spectra for the left-moving sector are stored in the exponent of the generating function as $E = L_0-\frac{c}{24}$, where $L_0$ have eigenvalues $h$ called the conformal weights or conformal dimensions \cite{orange_book} corresponding to the primary fields. There are different classes of RCFTs 
with different symmetry algebras and spectra.
We include two large classes of RCFTs: Wess-Zumino-Witten models characterized by affine Kac-Moody algebra and coset models.

Affine Kac-Moody algebra is the affine extension of simple Lie algebra. The spectra of affine Kac-Moody algebra CFTs can be realized through Sugawara construction of the energy-momentum tensor (see \ref{sec: sugawara} for a detailed review). Given an algebra $A$ with an arbitrary representation $r$, the conformal dimension and the central charge are defined as
\begin{equation} \label{eqn: h_and_c_formula}
h_r = \frac{C_r / \psi^2}{k + g} 
\hspace{0.5in}
c_A = \frac{k \, \mathrm{dim}\, A}{k + g}
\end{equation}
where $C_r$ is the quadratic Casimir, $k$ is known as levels or Coxeter numbers, $\psi$ is the highest root and $g$ is the dual Coxeter number \cite{carter2005lie, orange_book, applied_cft}. Meanwhile, the coset models are constructed from affine Kac-Moody algebra by quotienting it with another continuous group.

\begin{figure}
    \centering
    \includegraphics[width=.8\textwidth]{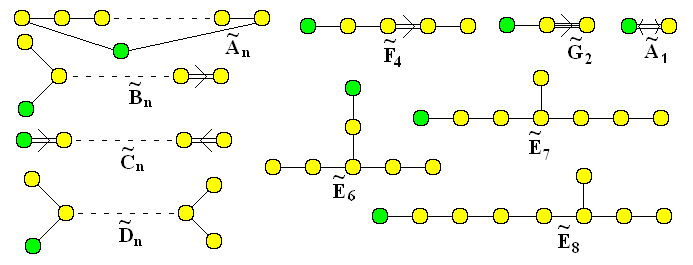}\hfill
    \caption{Extended Dynkin diagrams as classifications of affine Lie algebras.
    $\tilde{A}_n$ corresponds to $\widehat{sl}(n+1)$, which is a complexification of $\widehat{su}(n+1)$; $\tilde{B}_n$ corresponds to $\widehat{so}(2n+1)$; $\tilde{C}_n$ corresponds to $\widehat{sp}(2n)$; $\tilde{D}_n$ corresponds to $\widehat{so}(2n)$; and the rest are exceptional algebras. (source: Wiki)}
    \label{fig:dynkin}
\end{figure}

The classification of Kac-Moody algebra follows from
the extended Dynkin diagrams which 
classify simple Lie algebras.
In Fig \ref{fig:dynkin}, the $\widehat{su}(N)$ algebras \footnote{The hat notation $\widehat{su}(N)$ refers to the affine extension of simple Lie algebra $su(N)$. We apply this notation across other types of affine Lie algebra as well.} correspond to $\tilde{A}$-series where $N$ is any positive integer. There are multiple algebras other than the $\widehat{su}(N)$ as shown in Fig \ref{fig:dynkin}. In this paper, we include all series of Kac-Moody algebras shown in Fig \ref{fig:dynkin}. The complete Kac-Moody CFT library is presented in Section \ref{sec: data_generation}. We leave the detailed explanations of conformal dimensions and central charge for distinct Kac-Moody algebras to Appendix B.

\subsection{Tensor product theories}

According to the classification of modular invariant partition functions of two-dimensional conformal field theories \cite{cappelli2009ade}, there are three types of invariants: diagonal, non-diagonal, and exceptional invariant functions. We focus on tensor product of CFTs only from their diagonal invariants. In this way, the conformal dimensions and the central charges embedded in the exponent of the partition function are additive. 
The partition functions of the tensor product models are given by 
\begin{equation}
    Z_{i}^{(T)} = \prod_{n=1}^{L_i} Z_n = \prod_{n=1}^{L_i} \sum_{j\bar{j}}N_{j\bar{j}, n} \chi_{j,n}(q) \bar{\chi}_{\bar{j},n}(\bar{q})
\end{equation}
where $T$ refers to the tensor product theory and $n$ is the index that runs over
the number of constituent CFTs from $1,...,L_i$ and $i$ is the index for the tensor product theory. Here, we use $A_{n}$ to label the algebra associated with the constituent CFT such as $\widehat{su}(2)$. The central charge of $i$-th tensor product theory $c_i^{(T)}$ is the sum of central charges of $L_i$ constituent CFTs
\begin{equation}
    c_i^{(T)} = \sum_{n=1}^{L_i} c_{n}
\end{equation}
The algebra labels and central charges of constituent CFTs that form $i$-th tensor product theory are denoted by $A_{i,n}$ and $c_{i,n}$ separately. The conformal dimensions of $n$-th constituent CFT are defined as a set 
\begin{equation}
    \mathcal{H}_n = \Big\{ \ h_{n,l'_n} \ \Big| \ l'_n \in \mathrm{I}'_n \ \Big\}
\end{equation}
where $\mathrm{I}_n' = \{1,..., l_n'^{\mathrm{max}}\}$ is the set of allowed indices labeling elements of $\mathcal{H}_n$. The conformal dimensions of the $i$-th tensor product theory are defined as Minkowski sum of the sets $\mathcal{H}_1$, $\mathcal{H}_2$,..., $\mathcal{H}_{L_i}$:
\begin{equation}\label{eqn: h_tensor}
    \mathcal{H}^{(T)}_{i}= \Big\{ h_{i, l_i}^{(T)} = \sum_{n=1}^{L_i} h_{n, l'_n}  \ \Big| \ h_{n,l'_n} \in \mathcal{H}_n \ , \ l_i \in \mathrm{I}_i \Big\}
\end{equation}
where $\mathrm{I}_i = \{1, ..., l_i^{\mathrm{max}}\}$ is the set of allowed indices labeling elements of $\mathcal{H}_i^{(T)}$.



\section{RCFT Data Generation} \label{sec: data_generation}

We use holomorphic conformal dimensions $\mathcal{H}^{(T)}_i$ of the tensor product theories as the input for Transformers to predict the central charges of constituent CFTs $c_{i,n}$ and algebra label $A_{i,n}$ as output. Hence, we need to generate the RCFT data of the tensor product model by first computing the conformal dimensions for each constituent CFT and then adding them according to Eq.(\ref{eqn: h_tensor}).

We have a number of ways to make the RCFT datasets arbitrarily large, especially when taking tensor products. For a RCFT generated by the Kac-Moody algebra, we refer to Table \ref{fig:dynkin} 
for the 
nine distinct families of algebras. For each distinct algebra, we can take $N$ that's proportional to rank and the level $k$ to arbitrarily large numbers. The number of possible conformal dimensions in any constituent CFT can also be arbitrarily large. When taking tensor products of RCFTs, the number of constituent CFTs $L_i$ in a tensor product model can be arbitrarily large as well. Therefore, we impose additional constraints when generating the data. 




\begin{figure}
    \centering
    \includegraphics[width=0.48\linewidth]{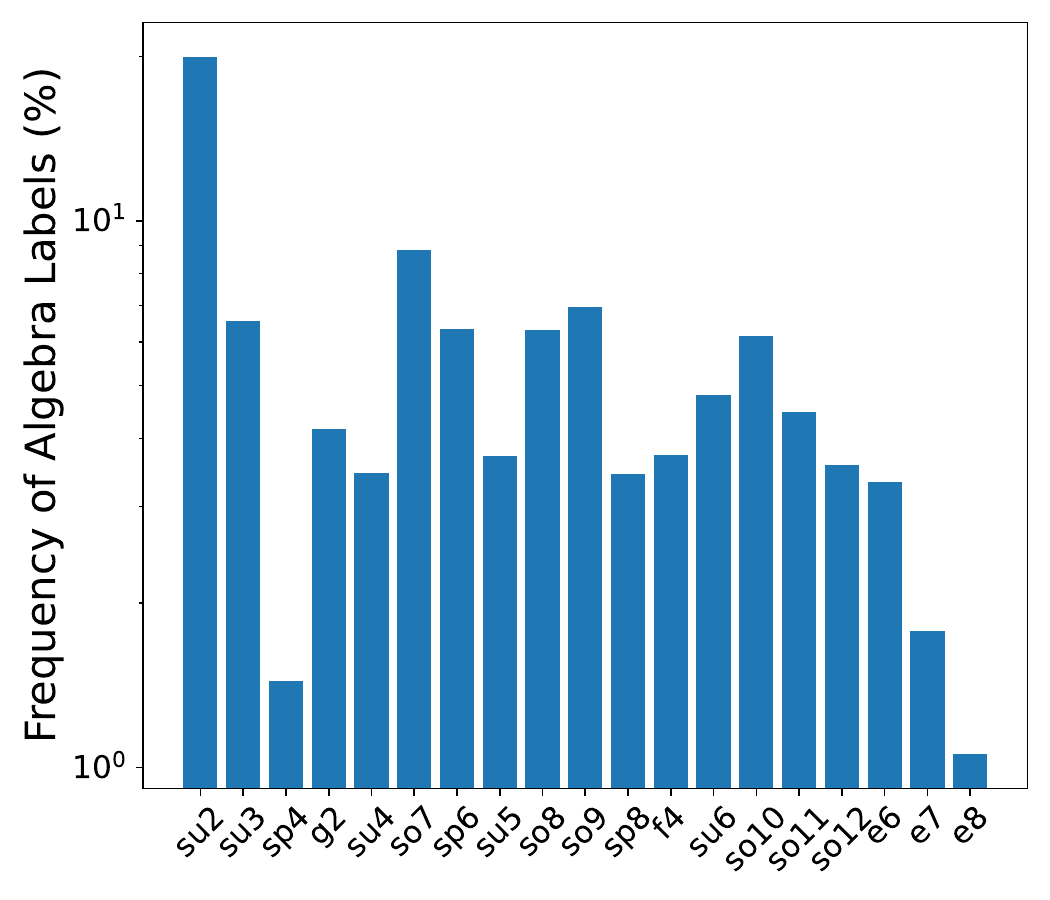}
    \includegraphics[width=0.48\linewidth]{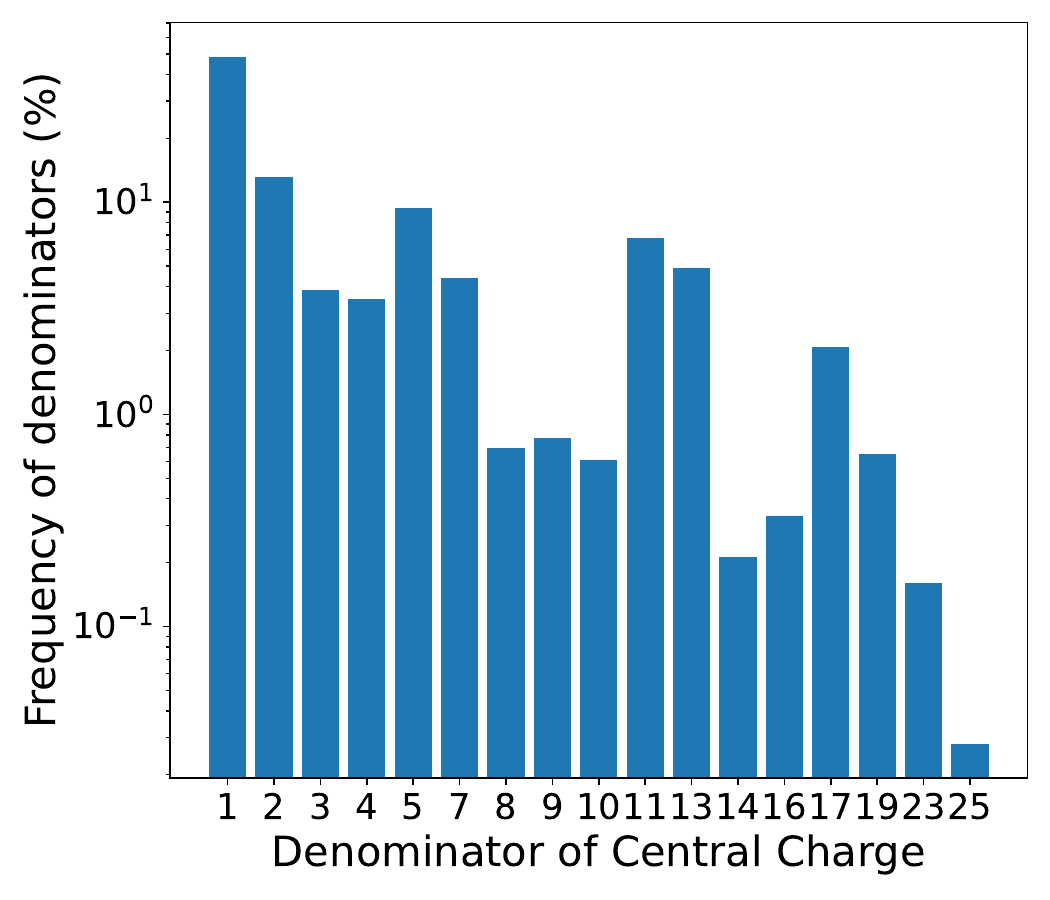}
    \caption{Algebra label and central charge distribution in the default train and test dataset ($c<50$) on a log-scale. Left: frequency of each algebra labels in the constituent theory; bottom: frequency of denominators of central charges in the constituent theory.}
    \label{fig: label_cc_freq_c50}
\end{figure}



We first constrain that $L_i \leq 5$ and we allow identical constituent CFTs to build up the same tensor product theory. 
Next, we restrict $N$ such that we make a RCFT library that contains theories generated by the following algebras: $\widehat{su}(2)$, $\widehat{su}(3)$, $\widehat{su}(4)$, $\widehat{su}(5)$, $\widehat{su}(6)$, $\widehat{so}(7)$, $\widehat{so}(9)$, $\widehat{so}(11)$, $\widehat{so}(8)$, $\widehat{so}(10)$, $\widehat{so}(12)$, $\widehat{sp}(4)$, $\widehat{sp}(6)$, $\widehat{sp}(8)$, $\widehat{E}_6$, $\widehat{E}_7$, $\widehat{E}_8$, $\widehat{F}_4$, and $\widehat{G}_2$. For each constituent CFT, we include theories with $k = 1,2,..,10$. In addition, we only include conformal dimensions smaller than $1$ for the constituent CFTs and we rank the conformal dimensions from lowest to highest by their numerical magnitudes. We denote these ordered sequences of conformal dimensions that are smaller than $1$ as the set $\tilde{\mathcal{H}}_n$ for the constituent CFTs. The total number of conformal dimensions included in each constituent CFT is restricted to be less than or equal to $6$. 
In the end, we have $144$ constituent CFTs with distinct lowest few conformal dimensions. Similarly, for the tensor product theory, we denote $\tilde{\mathcal{H}}^{(T)}_i$ as conformal dimensions that are smaller than $1$ and are ranked from lowest to highest by their numerical magnitudes. The total number of conformal dimensions included in each tensor product theory is restricted to be less than or equal to $51$. 

We observe that there are degenerate tensor product theories that have exactly identical conformal dimensions after truncation. Keeping all such tensor product theories in the training data will introduce an intrinsic ambiguity in the inverse problem: the same sequence of low-lying conformal dimensions corresponds to more than one tensor product theory. This intrinsic ambiguity due to the degeneracy complicates the evaluation as it can be conflated with the model's performance. Therefore, in order to isolate the model's performance, we  only retain one tensor product theory for each degenerate set. 
In the end, we construct around $3$ million tensor product theories with central charges no greater than $50$ from $144$ constituent CFTs. This is our default training and test dataset. We plot the histogram that gives the probability distribution of appearances for each algebra label in constituent CFTs in Fig \ref{fig: label_cc_freq_c50}. We notice that $\widehat{su}(2)$ dominate the dataset. In addition, we plot the probability distribution of appearances for denominators of central charges in constituent CFTs in Fig \ref{fig: label_cc_freq_c50} as well. 

In addition to the affine Kac-Moody algebras, we include the following coset constructions of Kac-Moody algebras (see \ref{sec: coset+tensor} for detailed reviews): Virasoro unitary minimal models, parafermion theory, $\mathcal{N}=1$ superconformal minimal models, and $\mathcal{N}=2$ superconformal minimal models. In Section \ref{sec: unseen_class}, we add these CFTs in the existing $144$ CFT library and make tensor product theories out of the new CFT library.




\section{Implementation} \label{sec: implement}

In this section, we discuss the setup of our task and the model implementation. We frame the inverse problem in CFT as a sequence-to-sequence task. For $i$-th tensor product theory, we provide as input their ordered sequences of conformal dimensions $\tilde{\mathcal{H}}_i^{(T)}$ to predict as output its $L_i$ constituent algebra labels $A_{i,n}$ and their corresponding central charge labels $c_{i,n}$ together as the set $\{(A_{i,n}, c_{i,n})\}_{n=1}^{L_i}$. The input conformal dimensions and the output central charges are expressed as reduced fractions. For an example of tensor product of two $\widehat{su}(2)$ theories with central charge equal to $1$, the training data looks like
\begin{verbatim}
              [0/1, 1/4, 1/2]   -->   [(su2, 1/1), (su2, 1/1)]    
\end{verbatim}
All integers are expressed in base 1000 and converted to tokens; the algebra labels (\texttt{su2}, \texttt{su3}, etc.) and the fraction symbol `$\texttt{/}$' are tokenized as well. Each algebra label is represented by a single token. For coset models, we tokenize them as a single token as well: $\texttt{minimal}$, $\texttt{parafermion}$, $\texttt{N1minimal}$, and $\texttt{N2minimal}$. We further separate the input and output sequence with a tab so the model distinguishes the feed-in data and the labels. The tensor product theories in our default dataset with input $\tilde{\mathcal{H}}_i^{(T)}$ and output $\{(A_{i,n}, c_{i,n})\}_{n=1}^{L_i}$ have central charges less than $50$ and Kac-Moody algebras as their labels, excluding the coset models. 

In all experiments, we use encoder-decoder Transformers built on an existing implementation used for theoretical physics work \cite{Cai:2024znx}, which contain a bidirectional Transformer encoder and an autoregressive Transformer decoder linked by a cross-attention mechanism \cite{attentionneed}. Both encoder and decoder have 2 layers, up to 8 attention heads. For all the models, the embedding dimension and Transformer dimension are the same (256). The result is evaluated using the cross-entropy loss function. Our models obtain the best performance with trainable parameters around 2.8 million for encoder and around 3.4 million for decoder. While many large language models \cite{brown2020languagemodelsfewshotlearners} have tens of billions of parameters, the size and scale of our models match those appeared in other AI for theoretical physics \cite{Cai:2024znx} and mathematics \cite{charton2022linearalgebratransformers} works. With minimum size and scale, our model achieves around $98\%$ accuracy score on in-domain data.

In this work, we define an epoch as a pass over $300,000$ sequences instead of a full pass over all training data ($3,000,000$ sequences), which makes the notion of epoch size more comparable between distinct experiments \cite{Cai:2024znx}. At the end of each epoch, the model is evaluated on a held-out test set. While there is no validation set as we are not performing hyperparameter tuning, the train-test set split always ensure that we have $10,000$ samples in the test set. The accuracy is defined as the number of correct output sequences $\{A_{i,n}, c_{i,n}\}_{n=1}^{L_i}$ predicted by models over the total number of output sequences ($10,000$) in the test set. In the default model, we use a learnable positional encoding in both the encoder and decoder. We use Adam optimizer \cite{adam} with learning rate of $10^{-4}$. No scheduler or warm-up implemented is in our task due to the small size of the models we use \cite{Cai:2024znx}. The models are trained on the following GPUs supported by Center for High Throughput Computing (CHTC) \cite{https://doi.org/10.21231/gnt1-hw21}: Tesla P100-PCIE-16GB, NVIDIA GeForce GTX 1080 Ti, NVIDIA GeForce GTX 2080 Ti, and NVIDIA A100-SXM4-40GB. Each run takes $61$ hours on average for $300$ epochs and $20$ hours for $100$ epochs.

As we mentioned above, we do not perform hyperparameter scanning since the goal of these experiments is not to optimize the final accuracy. Instead, our work aims to demonstrate that transformers are able to learn both algebraic and numerical structures of CFT, providing avenues for further studies in 2d CFTs.

\section{Results on the Inverse Problem}
\label{sec: base_model_training}

In the first experiment, we train encoder-decoder Transformers to predict the set $\{(A_{i,n}, c_{i,n})\}_{n=1}^{L_i}$ from ordered sequences of conformal dimensions $\tilde{\mathcal{H}}_i^{(T)}$ using cross-entropy loss. We restrict the central charge of the tensor product model to be less than $50$ in the training data and test on a held-out dataset unseen by the model containing central charges within the same range. The total number of samples in the training set is $3$ million whereas the number of samples in the validation set is $10,000$. 

\begin{table}[h]
\centering
\renewcommand{\arraystretch}{1.2}
\setlength{\tabcolsep}{8pt}
\begin{tabular}{@{} c c  @{\qquad} c c @{}}
\toprule
\multicolumn{2}{c}{\textbf{Training results (3M examples)}} \\
\midrule
\textbf{Metric} & \textbf{Accuracy (\%)} \\
\midrule
Permutation-invariant metric  & \textbf{97.7}  \\
Algebra label metric  & \textbf{97.9}       \\ 
Central charge metric  & \textbf{97.8}  \\
\bottomrule
\end{tabular}
\caption{Accuracies for tensor product theories with $c^{(T)}<50$.
Permutation-invariant accuracy measures the fraction of examples for which the set of predicted pairs of central charge and algebra labels match the ground-truth, regardless of the ordering of output pairs. Central charge accuracy measures the fraction of examples for which the predicted central charge matches the ground-truth. Algebra label accuracy measures the fraction of examples for which the predicted algebra label matches the ground-truth.}
\label{tab: acc_c<50}
\end{table}

We use three metrics to monitor the accuracy score of the model. 
As we mentioned above, distinct permutations of $\{(A_{i,n}, c_{i,n})\}_{n=1}^{L_i}$ result in the same tensor product theory. We thus use a metric called permutation-invariant accuracy. For each data sample or tensor product theory, the permutation-invariant metric measures the fraction of test examples for which the set of predicted pairs of central charge and algebra labels match the ground truth, regardless of the ordering of output pairs. For example, both $\{(\texttt{su2}, 1 \texttt{/} 1), (\texttt{su3}, 2 \texttt{/} 1)\}$ and $\{(\texttt{su3}, 2 \texttt{/}1), (\texttt{su2}, 1\texttt{/}1)\}$ are correct under permutation-invariant metric. Central charge accuracy measures the fraction of examples for which the predicted central charge matches the ground-truth. Algebra label accuracy measures the fraction of examples for which the predicted algebra label matches the ground-truth. For example, a predicted $\{(\texttt{su2}, 1  \texttt{/} 1), (\texttt{su3}, 2 \texttt{/} 1)\}$ will still be marked correctly on algebra label accuracy if the target sequence is $\{(\texttt{su2}, 1  \texttt{/} 1), (\texttt{su3}, 3 \texttt{/} 1)\}$, while this is not the case for central charge accuracy. Both algebra label and central charge accuracy metrics are permutation-invariant as well. Since a tensor-product theory is invariant under permutations of its constituent factors, we also tested an alternative setup in which the output pairs in the train and test datasets are randomly shuffled across the constituent CFTs for each tensor product theory. Under the permutation-invariant metrics, these results are similar to those obtained from our current setup in which the output pairs are not randomly shuffled. Hence, we report the results based on our current setup in the main text. 

We train Transformers for $300$ epochs and summarize the accuracy scores at the final epoch in Table \ref{tab: acc_c<50}. We see that the permutation-invariant accuracy reaches around $98\%$. This strongly indicates that the model is able to reconstruct the composition of tensor product theories but in various permutations that don't match the order of outputs in the target dataset. Thus, we conclude that the model successfully predicts the algebra labels and the central charges. Also, the algebra label accuracy and central charge accuracy are permutation-invariant. We see that the algebra label accuracy and central charge accuracy both align with the permutation-invariant accuracy, meaning that the model is learning the classification of algebra labels and regressing the central charge at the same rate. In conclusion, our Transformer model is able to reconstruct compositions of tensor products CFTs based on nearly identical low-lying conformal dimensions.

\subsection{Results for Algebra Label Prediction}


\begin{figure}
    \centering
    \includegraphics[width=1\linewidth]{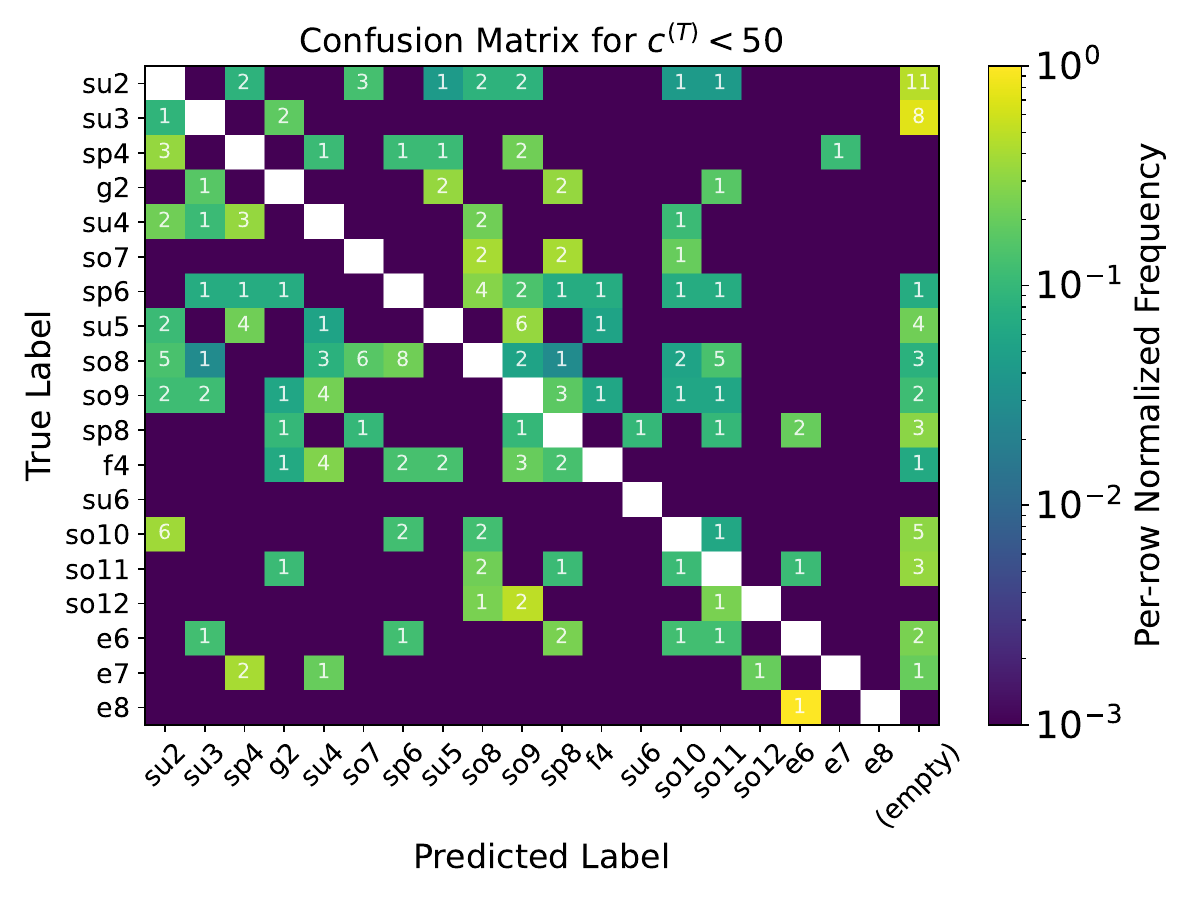}
    \caption{Confusion heatmap for algebra labels in tensor product theories with $c_i^{(T)}<50$ built from Kac-Moody CFT. The numbers for each row are normalized and then plotted on a log-scale. The number of incorrect predictions is shown on top of each color in the grid. The color also represent the distribution of incorrect predictions. The total number of incorrect predictions is $216$.}
    
    \label{fig: confusion_c50}
\end{figure}


We extend our analysis by looking at the prediction of algebra labels. Specifically, we look at the number of incorrect predictions for algebra labels. The complete information about incorrect predictions for each algebra labels is shown in the confusion heatmap in Fig \ref{fig: confusion_c50}. Each true algebra label can be mistaken as other algebra labels by Transformers. Each color on the heatmap represents the number of true labels being replaced by the predicted labels. 
In addition, the algebra labels on the axes are ordered by the magnitude of their rank. Starting from the smallest rank algebra $\texttt{su2}$ to the highest rank algebra $\texttt{e8}$. In Fig \ref{fig: label_cc_freq_c50}, $\texttt{su2}$ appears most frequently in the default dataset. If the number of times where $\texttt{su2}$ is mistaken as the true label when predicting other labels is the highest, it means that Transformers are learning based on frequency and memorization. However, in Fig \ref{fig: confusion_c50}, we observe that $\texttt{su2}$ is neither the dominant mistaken category nor the most incorrectly predicted label. Furthermore, the incorrect predictions distribute uniformly, indicating that there isn't an algebra label that is more likely to be the mistaken category by Transformers than other algebra labels. This shows that our Transformers are not making predictions based on frequency and memorization; rather, they are sensitive to some underlying algebraic structures from nearly-identical low-lying spectra of tensor product theories.

\subsection{Results for Central Charges Prediction}
Next, we examine the accuracy of central charge prediction for each class in default dataset. Based on Eq.(\ref{eqn: h_and_c_formula}),
learning the denominators of central charges is not a trivial memorization task because due to the combinatorial structure of conformal dimensions $\tilde{\mathcal{H}}^{(T)}_n$ of the tensor product theories, the denominators of $\tilde{\mathcal{H}}^{(T)}_n$ are different from that of $c_{i,n}$. Specifically, the denominator of $\tilde{\mathcal{H}}^{(T)}_n$ involves terms like $(k_1 + g_1)(k_2 + g_2)...(k_n + g_n)$. Furthermore, learning the numerators is nontrivial as there is no obvious relation between the quadratic Casimir $C_r$ and $k$ levels or $\mathrm{dim}  A$. In Table \ref{tab: acc_c<50}, our model achieves around $90\%$ accuracy score at the final epoch. Similar to what we have found in the previous subsection, these results show that our Transformers can extract the features from the low-lying conformal dimensions of tensor product models. These features include rich information about the symmetry algebra and the central charges of the tensor product theories.

\section{Generalization to Higher Central Charges} \label{sec: c-gen}

In the context of AdS/CFT, one focuses on theories with large central charges that approach semi-classical limits, suitable for finding their gravity duals. Motivated by this, we extend our task by including RCFTs with larger central charge. We explore if studying the algebraic and spectral properties of theories with lower central charges will enhance our understanding of those with higher central charges. 



\begin{figure}
    \centering
    \includegraphics[width=0.48\linewidth]{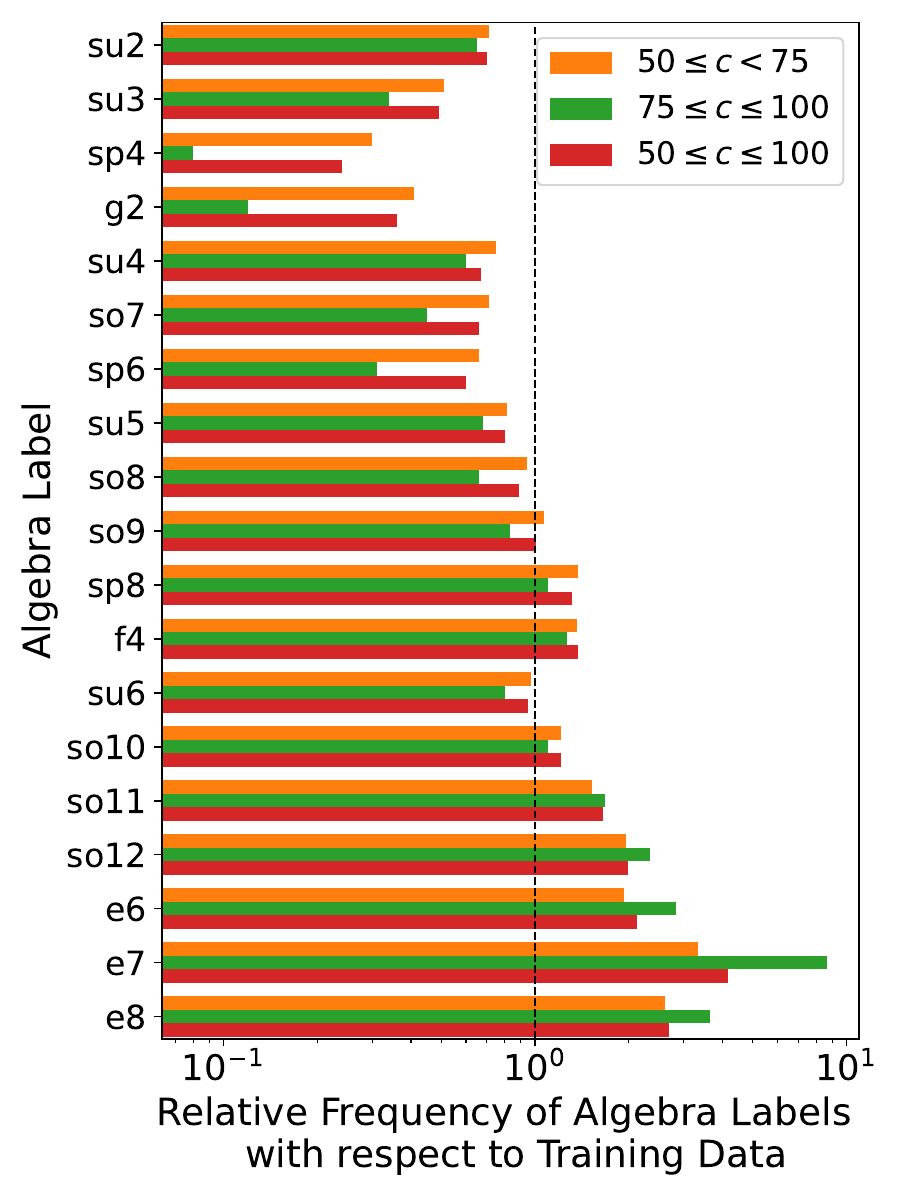}
    \includegraphics[width=0.48\linewidth]{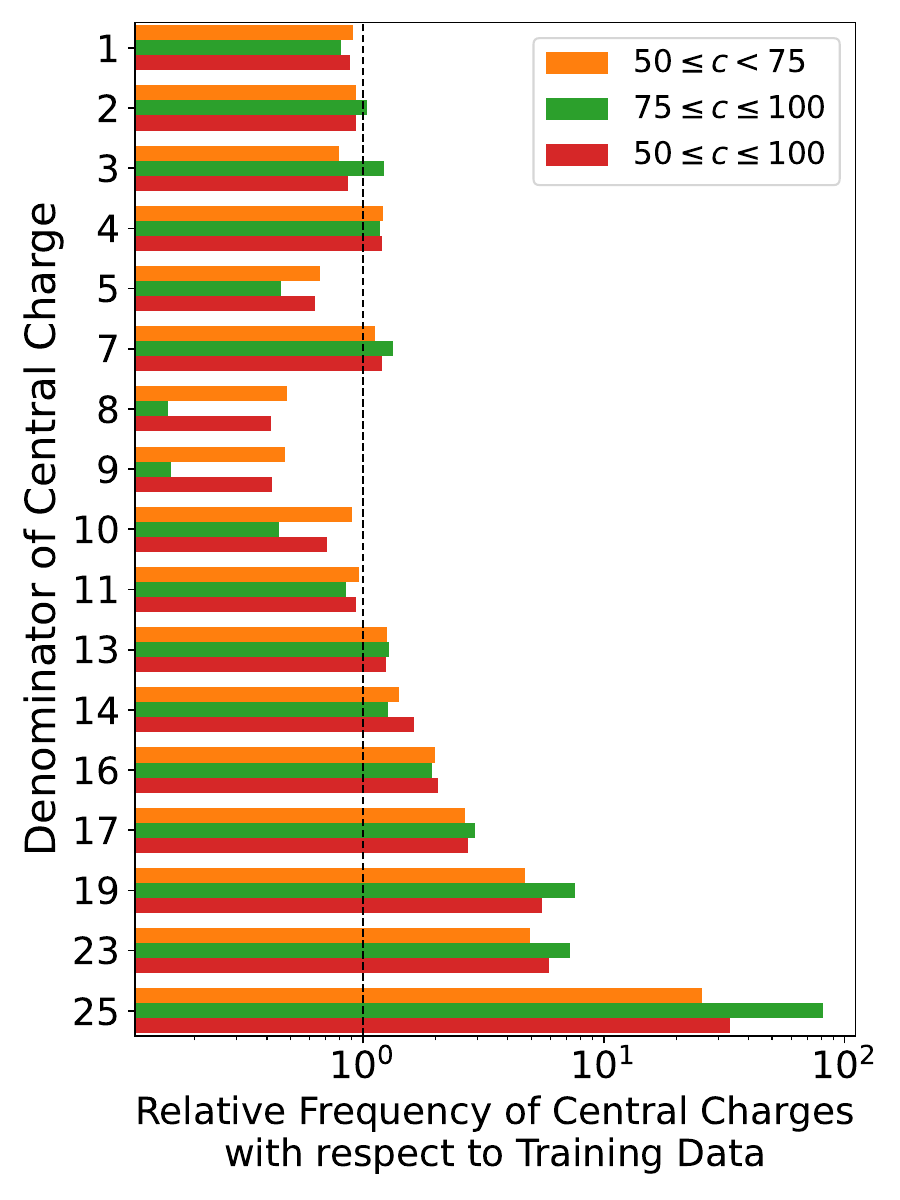}
    \caption{Relative frequency of algebra labels (left) and central charges (right) with respect to training data on a log scale. The different distributions between three colored bars represent domain shift.}
    \label{fig:freq_rel}
\end{figure}

In this section, we train the same encoder-decoder transformers on the default training dataset ($c^{(T)}<50$) alone and test on tensor product theories with central charges $50 \leq c^{(T)} \leq 100$ built from the same Kac-Moody constituent CFTs. We will show that the test data follows a different distribution from the training data: while the tensor product theories are built from the same constituent CFTs, higher central charges imply a higher number of constituent CFTs with larger central charge in each tensor product theory and more tensor product theories with a larger number of constituent CFTs. This is an example of domain shift. 

We will focus on three ranges of central charges: $50 \leq c^{(T)} < 75$, $75 \leq c^{(T)} \leq 100$, and $50 \leq c^{(T)} \le 100$. We plot their relative frequency of algebra labels and central charges with respect to training data ($c^{(T)}<50$) in Fig \ref{fig:freq_rel}. We observe that at larger $c^{(T)}$, higher rank algebras dominate the dataset while at smaller $c^{(T)}$, lower rank algebras dominate the dataset. Similarly, we observe that at larger $c^{(T)}$, central charges of constituent CFTs with larger denominators dominate the dataset while at smaller $c^{(T)}$, central charges of constituent CFTs with smaller denominators dominate the dataset. These are evidence of domain shift.

\subsection{``Zero-Shot" Generalization}
In this experiment, we use the same training data containing tensor product theories with $c^{(T)} < 50$ and test on completely unseen data (``zero-shot'') containing tensor product theories with $50 \leq c^{(T)} \leq 100$. The total number of distinct tensor product theories for training and test set is again $3,000,000$ and $10,000$. The accuracies are summarized in Table \ref{tab: acc_c50-100}. We see that the accuracy score on permutation-invariant metric reaches a $80\%$ scale. Even if it's a ``zero-shot'' experiment, the model exhibits relatively good performance on out-of-domain data. 

\begin{table}[h]
\centering
\renewcommand{\arraystretch}{1.2}
\setlength{\tabcolsep}{8pt}
\begin{tabular}{@{} c c  @{\qquad} c c @{}}
\toprule
\multicolumn{2}{c}{\textbf{Training results (3M examples)}} \\
\midrule
\textbf{Metric} & \textbf{Accuracy (\%)} \\
\midrule
Permutation-invariant metric  & \textbf{82.2}  \\
Algebra label metric  & \textbf{84.9}       \\ 
Central charge metric  & \textbf{82.2}  \\
\bottomrule
\end{tabular}
\caption{Accuracies for tensor product theories with $50 \leq c^{(T)} \leq 100$ built from Kac-Moody CFT trained on default training data with $c^{(T)}<50$.}
\label{tab: acc_c50-100}
\end{table}




We further analyze how much the domain shift affects the model's performance in this experiment. We made several subsequent experiments that test on different ranges of central charges while the CFT library remains fixed. Using the same default training model ($c^{(T)}<50$), we test on two additional sets of data with range of central charge: $50 \leq c^{(T)} < 75$ and $75 \leq c^{(T)} \leq 100$. The boundary for the first set is open as we don't want to include the same tensor product theories at $c^{(T)} =75$ in both sets. Also, we train the model on data with central charge $c^{(T)} \leq 100$ and test it on another two sets of data with central charge range: $c^{(T)} < 50$ and $50 \leq c^{(T)} \leq 100$.

In Fig \ref{fig:freq_rel}, we plot the frequency of algebra labels with different central charge ranges. We observe that there is a clear domain shift with respect to algebra labels as central charge varies. As central charges increase, the number of high rank algebras increases while the number of low rank algebras reduces. 



\begin{table}[h]
\centering
\renewcommand{\arraystretch}{1.2}
\setlength{\tabcolsep}{8pt}
\begin{tabular}{@{} l c  @{\qquad} l c @{}}
\toprule
\multicolumn{2}{c}{\textbf{Train on $c^{(T)} < 50$}} &
\multicolumn{2}{c}{\textbf{Train on $c^{(T)} \leq 100$}} \\
\cmidrule(lr){1-2}\cmidrule(lr){3-4}
\textbf{Test Data} & \textbf{Accuracy (\%)} &
\textbf{Test Data} & \textbf{Accuracy (\%)} \\
\midrule
$50 \leq c^{(T)} < 75$   & \textbf{91.3} & $ c^{(T)} < 50$    & \textbf{96.5} \\
$75 \leq c^{(T)} \le 100$  & 47.2 & $50 \leq c^{(T)} \le 100$  & \textbf{99.4} \\
$50 \le c^{(T)} \le 100$  & 84.2 &                     &       \\ 
\bottomrule
\end{tabular}
\caption{Permutation-invariant accuracies for different test data on two training ranges shown side-by-side.}
\label{tab: csweep}
\end{table}

In Table \ref{tab: csweep}, we report the model full accuracy for different experiments. We first train Transformers on default dataset containing tensor product theories within range $c^{(T)} < 50$ and test on three different sets of tensor product theories: $50 \leq c^{(T)} \leq 75$, $75 \leq c^{(T)} \leq 100$, and $50 \leq c^{(T)} \leq 100$. The domain shift phenomenon is shown in Fig \ref{fig:freq_rel}. We see that on the left, the accuracy score is dependent of the range of central charges. For test set which have $c^{(T)}$ close to the $c^{(T)} = 50$ threshold, Transformers perform well. However, for test set which have $c^{(T)}$ away from the $50$ threshold, Transformers show reduced performance. This training behavior is consistent with extrapolation decay in distribution-shift settings. Moreover, we trained the model on datasets with larger central charges $c^{(T)} \leq 100$ and test on two different datasets: $ c^{(T)} < 50$ and $50 \leq c^{(T)} \leq 100$. We see that the accuracy scores on two test sets are close to each other since the training datasets contain all of them. In conclusion, we observe domain shift caused by increasing the central charges. Generalizing to datasets close to the threshold ($c^{(T)}=50$) is easier than generalizing to datasets that are away from the threshold. 


\begin{figure}
    \centering
    \includegraphics[width=1\linewidth]{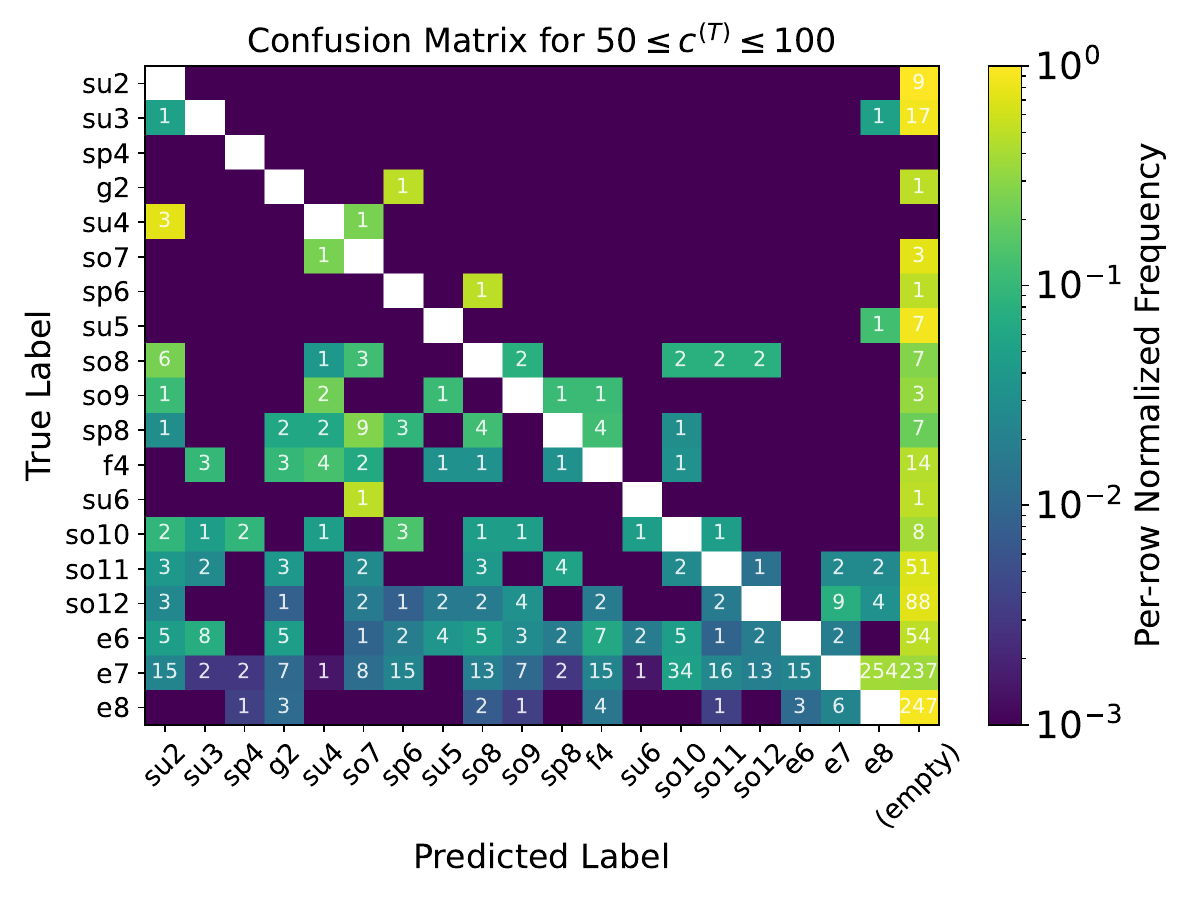}
    \caption{Confusion heatmap for algebra labels in tensor product theories with $50 \leq c^{(T)} \leq 100$ trained on tensor product theories with $c^{(T)} < 50$. The numbers for each row are normalized and then plotted on a log-scale. The number of incorrect predictions is shown on top of each color. The color also represent the distribution of incorrect predictions. The total number of incorrect predictions is $1396$.}
    \label{fig:confusion_c50-100}
\end{figure}

Regarding the ``zero-shot'' generalization, 
we plot the confusion heatmap for true and predicted algebra labels in Fig \ref{fig:confusion_c50-100}. 
In the heatmap, we see that there is not an obvious algebra label that is mistaken by Transformers more frequently than other labels similar to the phenomena we observed in Fig \ref{fig: confusion_c50}. In addition, we observe that the incorrect predictions distribute around the bottom half of the heatmap. This shows that the number of incorrect predictions for high rank algebras is higher than that of low rank algebras. This is in contrast with Fig \ref{fig: confusion_c50} where the distribution of incorrect predictions is relatively uniform across different algebra labels. This phenomenon shows that as central charge increases, the domain-shift is more prominent for higher rank algebras than that for lower rank algebras. Moreover, although higher rank algebras increase in their 
number,
they don't dominate over low rank algebras as shown in Fig \ref{fig:freq_rel}, thus this phenomenon is less likely to be caused by Transformers making predictions based on frequency.

\subsection{Generalization with Priming}

To further improve the model's capability of generalization, we apply a method introduced in Ref \cite{alfarano2024globallyapunovfunctionslongstanding}, called priming. Different from fine-tuning \cite{howard2018universallanguagemodelfinetuning} and transfer learning \cite{yosinski2014transferablefeaturesdeepneural}, priming means adding a small number of examples that follow the distribution of the test data to the training data and retraining the model. The added examples should be a tiny fraction of the training data. The small number of examples will enrich the training dataset with representative patterns that were previously unseen during training. This allows the model to adjust its internal representations to better align with the out-of-domain distribution. 
In this experiment, we add a tiny number of tensor product theories with $50 \leq c^{(T)} \leq 100$ to our default training data. We train the same Transformers on datasets with different number of added examples for $100$ epochs. The results are summarized in Table \ref{tab: cpriming}. The model accuracy increases monotonically as the number of added examples rises. We achieve around $97\%$ accuracy score when added examples are only $1\%$ of the default train data. 
Our experiment is performed on tensor product theories with central charges up to $100$. We can also extend the upper bound of the central charge to much higher values without massive generation of datasets and time, which is not the scope of this paper. Our findings demonstrate the ability of our Transformers trained on tensor product theories with lower central charges to generalize to theories with higher central charges by adding to the train dataset a small number of tensor product theories with larger central charges. 

\begin{table}
  \centering
  \renewcommand{\arraystretch}{1.4}
  \begin{tabular}{llll}
    \toprule
    Priming Data    & Amount of Priming &  Percentage of Priming  & Accuracy (\%) \\
    \cmidrule(r){1-4}
    None & 0 & 0 & 82.9\\
    \midrule
    $50 \leq c \leq 100$ & 30 & $0.001\%$ & 84.7 \\
    $50 \leq c \leq 100$ & 300 & $0.01\%$ & 85.4  \\
    $50 \leq c \leq 100$ & 3000 & $0.1\%$ & 93.3 \\
    $50 \leq c \leq 100$ & 30000 & $1\%$ & 97.0\\
    \bottomrule
  \end{tabular}
  \newline
  \caption{Priming results for central charge generalization. The percentage of priming is relative to the default training dataset.}
  \label{tab: cpriming}
\end{table}

\section{Generalization to Unseen Classes} \label{sec: unseen_class}

In the final experiment, we add a new class of CFT, the coset models, into our current Kac-Moody CFT library (see \ref{sec: coset} for details). The coset models include the Virasoro minimal models, the parafermionic theory, $\mathcal{N}=1$ superconformal theories, and $\mathcal{N}=2$ superconformal theories. A detailed derivation of the spectra of these theories can be found in 
\cite{orange_book, applied_cft, cappelli2009ade, Gepner_spacetime_susy, Gepner_exactly_solvable, Greene:1996cy}. The key difference between Kac-Moody CFTs and coset models is that they follow distinct algebraic construction. The coset CFTs are characterized by the W-algebras, which are cosets of Kac-Moody algebras. 


We add them in our default Kac-Moody CFT library so that the current CFT library contains $190$ theories in total. Then, we make tensor product theories based on the new library such that the tensor product theories contain a mixture of Kac-Moody CFTs and coset CFTs. We train the same Transformers on our default training dataset and test on tensor product theories that contain a mix of Kac-Moody CFTs and coset CFTs. In Table \ref{tab: coset_priming}, we report our findings. Initially, the Transformers with no priming data that contain a mix of Kac-Moody CFTs and coset CFTs added perform poorly. However, as we add more priming data, the model performance improves to $83.0\%$, which is an incredible jump compared to model performance with no priming data. This suggests that our Transformers can generalize to datasets that contain completely unseen constituent CFTs by adding a small number of out-of-domain examples. 

We observe that our Transformers' performance on generalizing to higher central charges is different from that on generalizing to tensor product theories with unseen constituents. In the latter case, Transformers require more priming data to reach higher accuracies, but the improvements are significant when we add a tiny more priming data to the training dataset.



\begin{table}
  \centering
  \renewcommand{\arraystretch}{1.4}
  \begin{tabular}{llll}
    \toprule
    Priming Data     & Amount of Priming & Percentage of Priming & Accuracy (\%) \\
    \cmidrule(r){1-4}
    None & 0  & 0 & 0\\
    \midrule
    Coset+Kac-Moody & 30 & 0.001\% & 1.2 \\
    Coset+Kac-Moody & 300 & 0.01\% & 17.5  \\
    Coset+Kac-Moody & 3000 & 0.1\% & 52.4 \\
    Coset+Kac-Moody & 30000 & 1\% & 83.0\\
    \bottomrule
  \end{tabular}
  \newline
  \caption{Priming results for unseen class generalization. The percentage of priming is relative to the default training dataset.}
  \label{tab: coset_priming}
\end{table}








\section{Conclusions}
In this study, we have shown that a Transformer model is able to reconstruct the compositions of tensor products of $2D$ rational conformal field theories based on their low-energy spectra with high accuracy.  We began with conformal field theories characterized by affine Kac-Moody algebras, known as Wess-Zumino-Witten realizations. We first examined tensor products of these CFTs with total central charge smaller than $50$, which is our default training dataset.
The model is effective at solving the inverse problem of predicting the algebra labels and central charges of the constituent CFTs based on low-lying conformal dimensions of tensor product theories. We further analyzed the prediction accuracy of the algebra labels and the central charges individually. 

Motivated by the semi-classical limits of AdS/CFT that involve large central charges, we study if Transformers can generalize to tensor product theories with larger central charges. 
We first quantified the domain shift that occurs as central charge $c^{(T)}$ increases. 
We show that our Transformers generalize better to datasets close to the threshold used for training $(c^{(T)}=50)$ than datasets that are away from that threshold. We also observe that as central charge $c^{(T)}$ increases, the domain-shift is more prominent for higher rank algebras than for lower rank algebras. 
Subsequently, we tested how the model generalizes to tensor product theories with higher central charge when adding a small amount of priming data to the training dataset. The model's performance improves as the amount of priming data increases. 

Finally, we test how the model generalizes to tensor product theories that contain an unseen class of CFTs (specifically, the coset CFTs). We tested how the model generalizes when adding a small amount of priming data to the training dataset. We observed that the model's performance improves as the amount  of priming data increases. However, we see that compared to generalization to higher central charges, the models  require more priming data to reach high accuracies for the unseen classes.

We note that the inverse map from truncated conformal dimensions to tensor product theories is not always unique. This degeneracy of tensor product theories sets an irreducible limit on the accuracy of any solution to the inverse problem, and that can easily be conflated with deficiencies in the model's performance. In this study, we retain only one tensor product theory per degenerate set in order to isolate the model's performance. Developing evaluation methods that explicitly resolve the ambiguity between the model's performance and the degeneracy of tensor product theories is an important next step and is left for future work. In addition, our method does not consider the full operator content of conformal field theories. While the operator spectrum fixes the scaling behavior of correlation functions, it does not determine their overall normalization, nor more refined dynamical data \cite{BELAVIN1984333, Polyakov:1974gs}. It is natural to ask whether machine learning techniques can be extended to learn these normalization factors and other dynamical variables and which representations of these dynamical variables \cite{Isono:2018rrb,Isono:2019wex,Bzowski:2019kwd,Bzowski:2020kfw, Gopakumar:2016wkt,Gopakumar:2016cpb}
are more suitable for machine learning applications. We leave a systematic exploration of these questions for future work.

\section{Acknowledgments}
A prior version of this work was presented in a non-archival form at the Machine Learning for the Physical Sciences \cite{cao2025learningCFTpysr}. This work utilizes the codes and packages in AI for scattering amplitudes project \cite{Cai:2024znx}. We would like to thank Mariel Pettee and Sanjit Shashi for helpful discussions. GS is partly supported by the U.S. Department of Energy, Office of Science, Office of High Energy Physics under Award Numbers DE-SC-0023719 and DE-SC-0017647. KC and HC are partially supported through the University of Wisconsin Madison, and the Wisconsin Alumni Research Foundation. GM and KC are supported by the U.S. Department of Energy (DOE) under Award No. DE-FOA-0002705, KA/OR55/22 (AIHEP).

\printbibliography[heading=bibintoc]
\newpage
\appendix

\section{Sugawara construction of WZW models}
\label{sec: sugawara}
Following from Section \ref{sec: basics}, we review the complete derivations and calculations of conformal dimensions and central charges. We start with Sugawara construction in the WZW models. The action of Wess-Zumino-Witten model can be written as the following:
\begin{equation}
    S = \frac{k}{16\pi} \int d^2x \ \mathrm{Tr}(\partial_{\mu}g)(\partial^{\mu}g^{-1}) + \frac{k}{24\pi}\int_B \mathrm{Tr}(g^{-1}dg)^3 
\end{equation}
where B is the three dimensional manifold and its boundary is the compactification of our original two-dimensional space. $g(x)$ is the matrix bosonic field living on the group manifold associated with the Lie algebra $A$. The equation of motion implies the conservation of Noether currents $J_{\mu} = g^{-1}\partial_{\mu}g$. We can write them in the complex coordinates and rescale by Coxeter number $k$ such that 
\begin{equation}
    \begin{split}
        J(z) &= -k J_z(z) = -k \partial_z g g^{-1} \\
        \bar{J}(\bar{z}) &= \  \ k \bar{J}_{\bar{z}}(\bar{z}) = \ \  k g^{-1} \partial_{\bar{z}}g
    \end{split}
\end{equation}
The currents can also be written as $J(z) = \sum_a J^a t^a$. Following from this, we have the operator product expansion of the currents, which is known as current algebra
\begin{equation}
    J^a(z) J^b(w) \sim \frac{k \delta_{ab}}{(z-w)^2} + \sum_{c} if_{abc} \frac{J^c(w)}{(z-w)}
\end{equation}
Here, the general group invariant realization of the energy-momentum tensor in terms of the current is 
\begin{equation}
    T(z) = \frac{1}{\beta} \sum_{a=1}^{\mathrm{dim} A}:J^a(z)J^a(z): 
\end{equation}
we fix the constant $\beta$ by requiring that $J^a(z)$'s transform as $(1,0)$ primary fields. We thus look at the operator product expansion $T(z)J^a(w)= \frac{1}{\beta}\sum_a :J^b(z)J^b(z):J^a(w)$. The singular terms are computed as follows:
\begin{equation}
    \begin{split}
        \wick{T\c(z)J^a\c(w)} &= \frac{1}{\beta}\sum_b    \wick{:J^b(z) \c J^b(z):J^a\c(w)}  \\
        &=\frac{1}{\beta} \sum_b \frac{1}{2\pi i} \oint_z \frac{dx}{x-z} \wick{[J^b\c(x)J^a\c(w)J^b(z) + J^b(x)J^b\c(z)J^a\c(w) ]} \\
        &=\frac{1}{\beta}\sum_b \frac{1}{2\pi i} \oint_z \frac{dx}{x-z} \Big\{\Big[ \frac{ k \delta_{ab}}{(w-x)^2} + \sum_cif_{bac}\frac{J^c(x)}{(w-x)} \Big]J^b(z)  \\
        &+ J^b(x)\Big[ \frac{k\delta_{ab}}{(z-w)^2} + \sum_c if_{bac}\frac{J^c(w)}{(z-w)}  \Big] \Big\} \\
    \end{split}
\end{equation}
we can simplify this expression by considering the following identities:
\begin{equation}
    - \sum_{b,c}f_{abc}f_{cbd} =  \sum_{b,c}f_{abc}f_{dbc} = 2g \delta_{ad}
\end{equation}
\begin{equation}
    \sum_{b,c}f_{abc}[(J^bJ^c)+ (J^cJ^b)] = 0
\end{equation}
We end up with 
\begin{equation}
    \begin{split}
        \wick{T\c(z)J^a\c(w)} &= \frac{1}{\beta} 2(k+g) \frac{J^a(w)}{(z-w)^2} \\
        &= \frac{1}{\beta}2(k+g) \Big\{ \frac{J^a(w)}{(z-w)^2} + \frac{\partial J^a(w)}{(z-w)} \Big\}
    \end{split}
\end{equation}
Thus, for $T$ to be a genuine energy-momentum tensor, the coefficient must be 1 based on the primary field criterion, which gives us
\begin{equation}
    T(z) = \frac{1/2}{k + g}\sum_{a=1}^{\mathrm{dim}  A}:J^a(z)J^a(z):
\end{equation}
Now, we can compute the operator expansion of two energy-momentum tensor in its singular terms:
\begin{equation}
    \begin{split}
        \wick{T\c(z)T\c(w)} &= \frac{1/2}{k+g} \frac{1}{2\pi i}\oint_w \frac{dx}{x-w} \sum_a \Big\{ \wick{T\c(z) J^a\c(x) J^a(w)} + \wick{J^a(x) T\c(z)J^a\c(w)} \Big\} \\
        &= \frac{1/2}{k+g} \frac{1}{2\pi i}\oint_w \frac{dx}{x-w} \sum_a \Big\{ \frac{J^a(w) J^a(z)}{(z-x)^2} + \frac{\partial J^a(w) (z-w) J^a(z)}{(z-x)^2} 
        + \frac{J^a(x) J^a(w)}{(z-w)^2} \\
        &+ \frac{J^a(x) \partial J^a(w)}{(z-w)} \Big\} \\
        &= \frac{1/2}{k+g} \frac{1}{2\pi i}\oint_w \frac{dx}{x-w} \sum_a \Big\{ \frac{J^a(x)J^a(w)}{(z-x)^2} - \frac{2k\delta_{aa}}{(z-x)(x-w)^3} + \frac{J^a(x)J^a(w)}{(z-w)^2} \\
        &+ \frac{J^a(x) \partial J^a(w)}{(z-w)} \Big\} \\
        &= \frac{k \mathrm{dim} \ A}{k + g} \frac{1/2}{(z-w)^4} + \frac{2T(w)}{(z-w)^2} + \frac{\partial T(w)}{(z-w)}
    \end{split}
\end{equation}
where we used Cauchy integral $f^{(n)}(a) = \frac{n!}{2\pi i} \oint \frac{f(z)}{(z-a)^{n+1}}dz$, $\delta_{aa}= \mathrm{dim} \ A$, and the following OPE:
\begin{equation}
    \partial J^a(z) J^b(w) \sim \frac{-2k\delta_{aa}}{(z-w)^3} - \sum_c i f_{abc} \frac{J^c(w)}{(z-w)^2}
\end{equation}
Hence, the coefficient of the last term is central charge $c$. Now we look at the operator product expansion between the current operator and the primary field $\varphi^l(r)$
\begin{equation}
    J^a(z)\varphi_{(r)}(w) \sim \frac{t^a_{(r)}}{z-w}\varphi_{(r)}(w) + ...
\end{equation}
Here $\varphi_{(r)}$ transforms as some representation $(r)$ of $G$. These primary fields create states called highest weight states:
\begin{equation} \label{highest weight state}
    \ket{(r)} = \varphi_{(r)}(0)\ket{0} 
\end{equation}
\begin{equation}
    J^a_0\ket{(r)} = t^a_{(r)}\ket{(r)}
\end{equation}
where $t^a_{(r)}$ is the representation matrices for G in the representation $(r)$. The generator of the conformal algebra is defined as 
\begin{equation}
    L_n = \frac{1/2}{k + g}\sum_a \sum_m :J_m^a J_{n-m}^a:
\end{equation}
If we apply the generator $L_0$ where $n=0$ on a state with an arbitrary representation, we get the following:
\begin{equation}
    L_0\ket{(r)} = \frac{1/2}{\tilde{k} + C_A/2}\sum_{a=1}^{|G|}:J_m^aJ_{-m}^a:\ket{(r)} = \frac{C_r/2}{\tilde{k} + C_A/2}\ket{(r)}
\end{equation}
where $C_r$ is the quadratic Casimir of the representation.

\section{Properties of Simple Lie Algebras and Spectra of CFT} \label{sec: appendix_spectra}
In this section, we review the algebraic properties of Lie algebras and summarize the complete derivation of conformal dimensions and central charges. We start with conformal dimensions defined as 
\begin{equation}
    h_r = \frac{C_r/2}{k + g}
\end{equation}
The quadratic Casimir operator $Q_r$ has the eigenvalue $C_r$ with respect to the highest weight state $\ket{\lambda}$
\begin{equation}
    Q_r \ket{\lambda} = (\lambda, \lambda+2\rho)\ket{\lambda} = C_r \ket{\lambda}
\end{equation}
in terms of the Weyl vector $\rho = \sum_i \omega_i$ where $\omega_i$ are called fundamental weights. The highest weight $\lambda$ can be written as a sum over the fundamental weights with coefficients called the Dynkin label $\lambda_i$
\begin{equation}
    \lambda = \sum_i^r \lambda_i \omega_i
    \hspace{0.2in}
    \lambda = (\lambda_1, ..., \lambda_r)
\end{equation}
In conformal field theory, the highest weight state corresponds to the primary fields acting on the vacuum state, which is $\ket{(r)}$ and we can check $L_n \ket{(r)} = 0$ for $n > 0$. Following from this, we can compute the conformal dimensions as 
\begin{equation}
    \begin{split}
        h_r &= \frac{(\lambda, \lambda + 2\rho)}{2(k + g)} \\
        &= \frac{(\lambda, \lambda) + 2(\lambda, \rho)}{2(k + g)} \\
    \end{split}
\end{equation}
By the integrability condition $(\lambda, \theta) \leq k$ where $\theta$ is the highest root, we can find the proper fundamental weights for the Dynkin labels. Then, we compute the conformal dimensions for each available Dynkin label. 
\begin{equation}
    \begin{split}
        h_{r,i} &= \frac{(\omega_i, \omega_i) + 2(\omega_i, \sum_j^r \omega_j)}{2(k +g)} \\
    \end{split}
\end{equation}
The fundamental weights can be extracted from the quadratic form matrix $F_{ij} = (\omega_i, \omega_j)$, which is proportional to the inverse of the Cartan matrix. The derivation of central charge is much easier. We start with the expression
\begin{equation}
    c_A = \frac{k \ \mathrm{dim} \ A}{k + g}
\end{equation}
where $k$ is Coxeter number or level, the same number that appears in the action of WZW model. $g$ is the dual Coxeter number and $\mathrm{dim}\ A$ is the dimension of the algebra $A$. Below, we summarize all the algebraic properties of simple Lie algebras used to calculate the spectra.
\begin{center}
\begin{tabular}{ |p{2.3cm}||p{2.5cm}|p{2.5cm}|p{2.5cm}|p{2.5cm}|  }
 \hline
 \multicolumn{5}{|c|}{Simple Lie Algebras $A$} \\
 \hline
 Properties & $su(r+1)$ & $so(2r+1)$ & $sp(2r)$ & $so(2r)$\\
 \hline
 series &  $\mathrm{A}_{r\geq2}$  & $\mathrm{B}_{r\geq3}$ & $\mathrm{C}_{r\geq2}$ & $\mathrm{D}_{r\geq4}$ \\
 $\mathrm{dim}  A$   & $r^2 + 2r$ & $2r^2+r$ & $2r^r + r$ & $2r^2-r$ \\
 $g$ &   $r+1$  & $2r-1$  & $r+1$ & $2r-2$ \\
 $\theta$ & (1,0,...,1) & (0,1,...,0) &  (2,0,...,0) & (0,1,...,0)\\
 \hline
\end{tabular}
\end{center}
\begin{center}
\begin{tabular}{ |p{2cm}||p{2cm}|p{2cm}|p{2cm}|p{2cm}|p{2cm}| }
 \hline
 \multicolumn{6}{|c|}{Exceptional Simple Lie Algebras $A$} \\
 \hline
 Properties & $E_6$ & $E_7$ & $E_8$ & $F_4$ & $G_2$ \\
 \hline
 $\mathrm{dim}  A$   & $78$ & $133$ & $248$ & $52$ & $14$ \\
 $g$ &   $12$  & $18$  & $30$ & $9$ &  $4$\\
 $\theta$ & (0,0,...,1) & (1,0,...,0) &  (1,0,...,0) & (1,0,0,0) & (1,0) \\
 \hline
\end{tabular}
\end{center}
The level $k$ is any positive integers starting from $1$. 
We also include the quadratic form matrices for all simple Lie algebras involved. For $su(r+1)$ algebra, we have 
\begin{equation}
\frac{1}{r+1}
\begin{pmatrix}
r & r-1 & r-2 & \cdots & 2 & 1 \\
r-1 & 2(r-1) & 2(r-2) & \cdots & 4 & 2 \\
r-2 & 2(r-2) & 3(r-2) & \cdots & 6 & 3 \\
\vdots & \vdots & \vdots & \ddots & \vdots & \vdots \\
2 & 4 & 2(r-1) & \cdots & 2(r-1) & r-1 \\
1 & 2 & r-1 & \cdots & r-1 & r
\end{pmatrix}
\end{equation}
For $so(2r+1)$, we have 
\begin{equation}
    \frac{1}{2}
    \begin{pmatrix}
        2 & 2 & 2 & \cdots & 2 & 1 \\
        2 & 4 & 4 & \cdots & 4 & 2 \\
        2 & 4 & 6 & \cdots & 6 & 3 \\
        \vdots & \vdots & \vdots & \ddots & \vdots & \vdots \\
        2 & 4 & 6 & \cdots & 2(r-1) & r-1 \\
        1 & 2 & 3 & \cdots & r-1 & r/2
    \end{pmatrix}
\end{equation}
For $sp(2r)$, we have 
\begin{equation}
    \frac{1}{2}
    \begin{pmatrix}
        1 & 1 & 1 & \cdots & 1 & 1\\
        1 & 2 & 2 & \cdots & 2 & 2\\
        1 & 2 & 3 & \cdots & 3 & 3\\
        \vdots & \vdots & \vdots & \ddots & \vdots & \vdots \\
        1 & 2 & 3 & \cdots & r-1 & r-1 \\
        1 & 2 & 3 & \cdots & r-1 & r
    \end{pmatrix}
\end{equation}
For $so(2r)$, we have 
\begin{equation}
    \frac{1}{2}
    \begin{pmatrix}
        2 & 2 & 2 & \cdots & 2 & 1 & 1\\
        2 & 4 & 4 & \cdots & 4 & 2 & 2 \\
        2 & 4 & 6 & \cdots & 6 & 3 & 3 \\
        \vdots & \vdots & \vdots & \ddots & \vdots & \vdots & \vdots \\
        2 & 4 & 6 & \cdots & 2(r-2) & r-2 & r-2 \\
        1 & 2 & 3 & \cdots & r-2 & r/2 & (r-2)/2 \\
        1 & 2 & 3 & \cdots & r-2 & (r-2)/2 & r/2
    \end{pmatrix}
\end{equation}
For $E_6$, we have 
\begin{equation}
    \frac{1}{3}
    \begin{pmatrix}
        4 & 5 & 6 & 4 & 2 & 3\\
        5 & 10 & 12 & 8 & 4 & 6 \\
        6 & 12 & 18 & 12 & 6 & 9\\
        4 & 8 & 12 & 10 & 5 & 6 \\
        2 & 4 & 6 & 5 & 4 & 3 \\
        3 & 6 & 9 & 6 & 3 & 6 
    \end{pmatrix}
\end{equation}
For $E_7$, we have 
\begin{equation}
    \frac{1}{2}
    \begin{pmatrix}
        4 & 6 & 8 & 6 & 4 & 2 & 4 \\
        6 & 12 & 16 & 12 & 8 & 4 & 8 \\
        8 & 16 & 24 & 18 & 12 & 6 & 12 \\
        6 & 12 & 18 & 15 & 10 & 5 & 9 \\
        4 & 8 & 12 & 10 & 8 & 4 & 6\\
        2 & 4 & 6 & 5 & 4 & 3 & 3\\
        4 & 8 & 12 & 9 & 6 & 3 & 7 \\
    \end{pmatrix}
\end{equation}
For $E_8$, we have 
\begin{equation}
    \begin{pmatrix}
        2 & 3 & 4 & 5 & 6 & 4 & 2 & 3\\
        3 & 6 & 8 & 10 & 12 & 8 & 4 & 6 \\
        4 & 8 & 12 & 15 & 18 & 12 & 6 & 9\\
        5 & 10 & 15 & 20 & 24 & 16 & 8 & 12\\
        6 & 12 & 18 & 24 & 30 & 20 & 10 & 15 \\
        4 & 8 & 12 & 16 & 20 & 14 & 7 & 10\\
        2 & 4 & 6 & 8 & 10 & 7 & 4 & 5\\
        3 & 6 & 9 & 12 & 15 & 10 & 5 & 8
    \end{pmatrix}
\end{equation}
For $F_4$, we have 
\begin{equation}
    \begin{pmatrix}
        2 & 3 & 2 & 1\\
        3 & 6 & 4 & 2\\
        2 & 4 & 3 & 3/2\\
        1 & 2 & 3/2 & 1 \\
    \end{pmatrix}
\end{equation}
For $G_2$, we have
\begin{equation}
    \begin{pmatrix}
        2 & 1 \\
        1 & 2/3 
    \end{pmatrix}
\end{equation}

\section{Coset Construction and Tensor Product Models} \label{sec: coset+tensor}
In this section, we introduce two ways of creating new classes of CFTs from manipulation of the known theories at hand: coset construction and tensor products. Both methods can greatly enlarge the conformal theory landscape. 

\subsection{coset constructions}\label{sec: coset}
We introduce the coset construction method and review the spectra and algebraic properties of coset models we included. 

Consider an affine Lie algebra $A$ and let $H$ be a subalgebra of $A$. 
The central charge is the difference of the central charges of the constituent models:
\begin{equation}
    c_{A/H} = c_A - c_H = \frac{k_A \mathrm{dim} \ A}{k_A + g} - \frac{k_H |H|}{k_H + \tilde{h}_H}
\end{equation}
We obtain the coset description of unitary minimal models as the following coset:
\begin{equation}
    \frac{\widehat{su}(2)_k \oplus \widehat{su}(2)_1}{\widehat{su}(2)_{k+1}}
\end{equation}
which is the $W(N)$-algebra with $N=2$. The conformal dimensions and the central charges are given by 
\begin{equation}
    h_{r,s} = \frac{((k+3)r-(k+2)s)^2 - 1}{4(k+3)(k+2)}~,
    \hspace{0.5cm}
    c = 1-\frac{6}{(k+2)(k+3)}
\end{equation}
The $(r,s)$ pairs represent the quantum number for energy levels and satisfy $1 \leq r \leq k+1$ and $1\leq s \leq k+2$. When $k=1$, the minimal models returns to the two-dimensional critical Ising model in statistical mechanics. Another interesting coset is the parafermionic model with
\begin{equation}
    \frac{\widehat{su}(2)_k}{\widehat{u}(1)_k}
\end{equation}
where $\widehat{u}(1)$ corresponds to a free boson theory living on a circle of radius $\sqrt{2k}$. The conformal dimensions and central charges are given by
\begin{equation}
\label{eqn: para_h_c}
    h_{l,m} = \frac{l(l+2)}{4(k+2)} - \frac{m^2}{4k}~,
    \hspace{0.5 cm}
    c = \frac{3k}{k+2} - 1 = \frac{2(k-1)}{k+2}
\end{equation}
where $l = 0,1,...,k$ and $l+m = 0 \ \mathrm{mod}\ 2$ with $m \ \mathrm{mod} \ 2k$.


We also include coset models that admit supersymmetry. Two well-known two dimensional conformal field theories are $\mathcal{N}=1$ superconformal minimal models and $\mathcal{N}=2$ superconformal minimal models. The coset realization of $\mathcal{N}=1$ superconformal minimal models resemble that of unitary minimal models:
\begin{equation}
    \frac{\widehat{su}(2)_k \oplus \widehat{su}(2)_2}{\widehat{su}(2)_{k+2}}
\end{equation}
The conformal dimensions in the Neveu-Schwarz (NS) sector and central charges are given by
\begin{equation}
    h^{\mathrm{NS}}_{r,s}= \frac{\big[(m+2)r - m s\big]^2 - 4}{8m(m+2)} + \frac{1}{32}(1-(-1)^{r-s})
    \hspace{0.5 cm}
    c = \frac{3}{2}\left(1 - \frac{8}{m(m+2)}\right)
\end{equation}
where $1 \le r \le m-1,\; 1 \le s \le m+1,\; r-s \equiv 0 \pmod{2}$. Notice that $m$ here is not the same as $m$ in Eq \ref{eqn: para_h_c}. Finally, we introduce the $\mathcal{N}=2$ superconformal minimal models
\begin{equation}
    \frac{\widehat{su}(2)_k \oplus \widehat{u}(1)_2}{\widehat{u}(1)_{k+2}}
\end{equation}
The conformal dimensions in the NS sector and central charges are given by
\begin{equation} \label{h_N2}
    h^{\mathrm{NS}}_{\ell,m,s}=\frac{\ell(\ell+2) - m^2}{4(k+2)}+\frac{s^2}{8} + \mathrm{integer}
    \hspace{0.5 cm}
    c = \frac{3k}{k+2}
\end{equation}
where $l = 0, 1,...,k$, $m \in \mathbb{Z}_{2(k+2)}$, and $s \in 2\mathbb{Z}$ in the NS sector. The labels have to satisfy $l+m+s \in 2\mathbb{Z}$ and there is an identification between primary fields with labels $(l,m,s) \sim (k-l, m+k+2, s+2)$. With the help of this identification, we can bring the triplet of parameters $(l,m,s)$ to the range $0 \leq |m-s| \leq l$ and $s \in \{-1, 0,1,2\}$ such that the integers in Eq \ref{h_N2} can be zero. The only exceptions in the NS sectors are representations with labels $(l,-l,\ 2)$.

\subsection{tensor product models} \label{sec: tensor product}
\begin{table}
\caption{\label{jlab1} List of critical dimensions and central charges for distinct string theories. For superstring theory, the central charge includes both bosonic ($c=1$) and fermionic ($c=\frac{1}{2}$) strings. For heterotic string, the left- and right-moving sector has different central charges.}
\footnotesize
\begin{tabular}{@{}llll}
\br
Theory Type & Critical Dimension & Spacetime Sector ($c$) & Internal Sector ($c_{int}$)\\
\mr
Bosonic String & $26$ & $c=d$ & $c_{int} = 26 - d$ \\ 
Type II Superstring & $10$ & $c = d + \frac{1}{2}d = \frac{3}{2}d$ & $c_{int} = 15 - \frac{3}{2}d$ \\
Heterotic String & 26 (Left), 10 (Right) & $c_L = d$, $c_R = \frac{3}{2}d$ & $c_{int,L} = 26 - d$, $c_{int,R} = 15 - \frac{3}{2}d$ \\
\br
\end{tabular}\\
$^{a}$d is the spacetime dimension of the theory. In reality, $d=4$. 
\label{tab: central charge table}
\end{table}
\normalsize

We now review the tensor product theories and our motivations of using this to expand the CFT landscape.

The tensor product of CFTs appears frequently in the context of string compactifications. Quantum consistency of the worldsheet theory, 
if realized by free fields, requires that the total number of spacetime dimensions should be $26$ for the bosonic string and $10$ for the superstring.
Therefore, to obtain realistic particle physics like the Standard Model in our 4d spacetime dimensions, the extra dimensions are compactified. More generally, the extra dimensions are realized as an internal CFT with an appropriate central charge (see \cite{Marchesano:2024gul} for a review).
Tensor product of RCFTs can be good candidate theories that describe the worldsheet of the internal sector as long as their central charges sum up to the correct values as shown in Table \ref{tab: central charge table}.
\begin{equation}
    \mathcal{C} = \mathcal{C}_1 \otimes \mathcal{C}_2 \otimes ...
\end{equation}
The constituent CFTs $\mathcal{C}_i$ of the bosonic string can be any RCFTs with an extended Kac-Moody algebra. For superstring, we have to include spacetime supersymmetry to allow the existence of fermions. One explicit example is the Gepner model \cite{Gepner_exactly_solvable, Gepner_spacetime_susy}. Gepner found
that if one tensor product five $\mathcal{N}=2$ minimal models at level $3$ orbifolding odd $U(1)$ charges such that $c_{internal}=9$, the worldsheet theory
is exactly solvable and the massless spacetime spectra and symmetries coincide with the 
quintic Calabi-Yau space
given by the vanishing locus of $z_1^5 + z_2^5 + z_3^5 + z_4^5 + z_5^5$ in $\mathbb{C}P^4$ \cite{Gepner_exactly_solvable, Gepner_spacetime_susy}. 

\section{Learning Dynamics of Transformers} \label{sec: learning_dynamics}
\begin{figure}
    \centering
    \includegraphics[width=.49\textwidth]{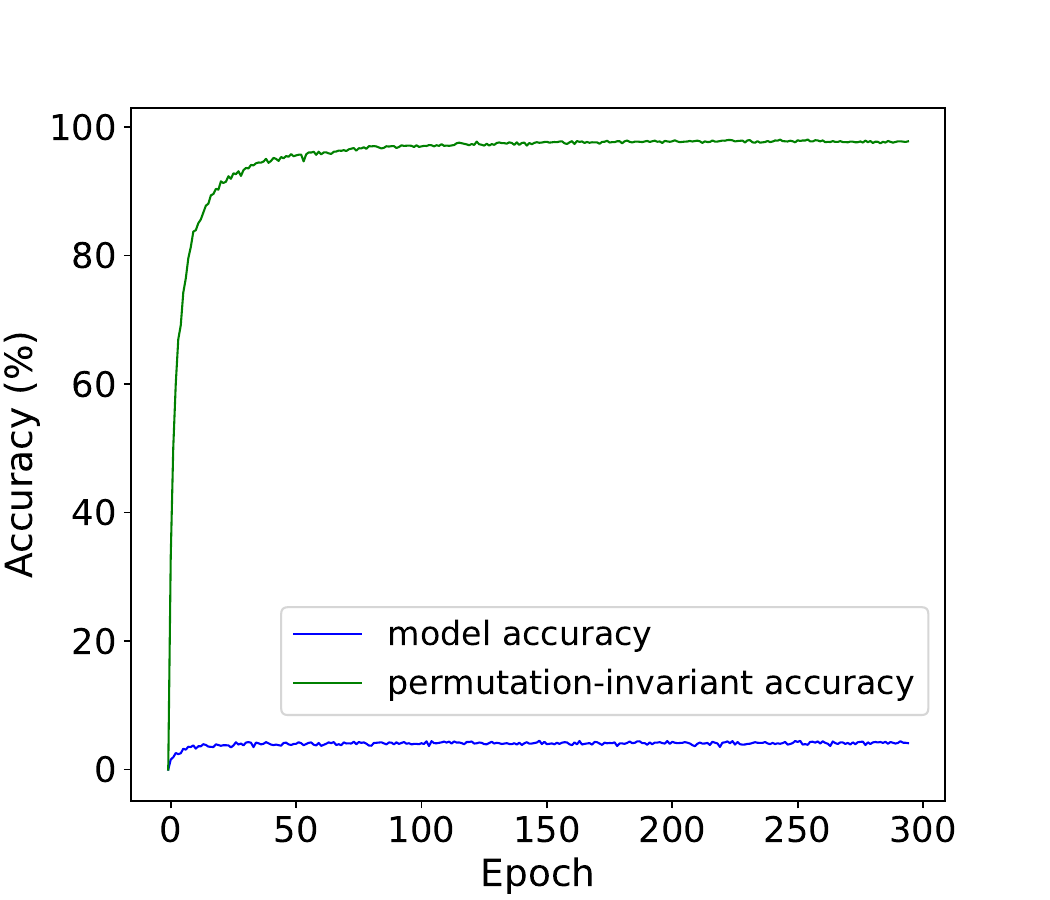}
    \includegraphics[width=.49\textwidth]{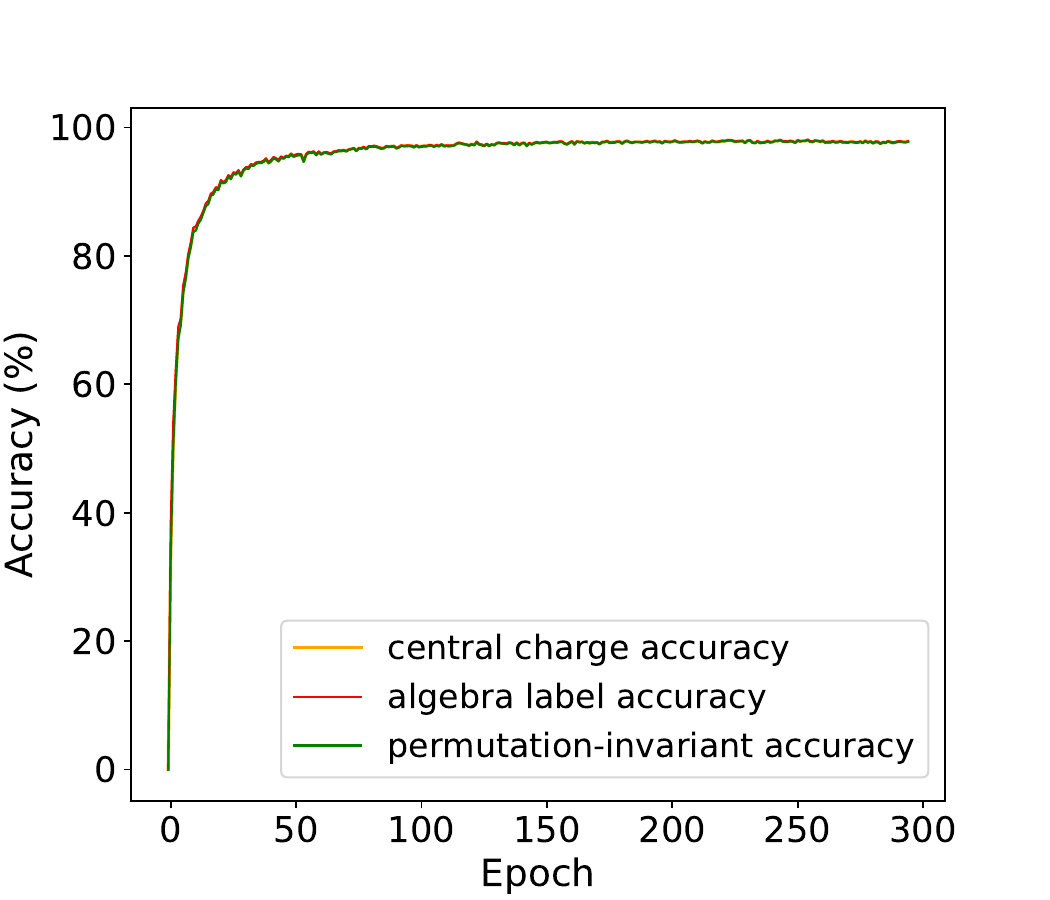}
    \caption{Accuracy Curves for the base model training: tensor products of WZW models generated by the KM algebras.}
    \label{fig:cft_KM_c50_base_acc}
\end{figure}

\begin{figure}
    \centering
    \includegraphics[width=0.48\linewidth]{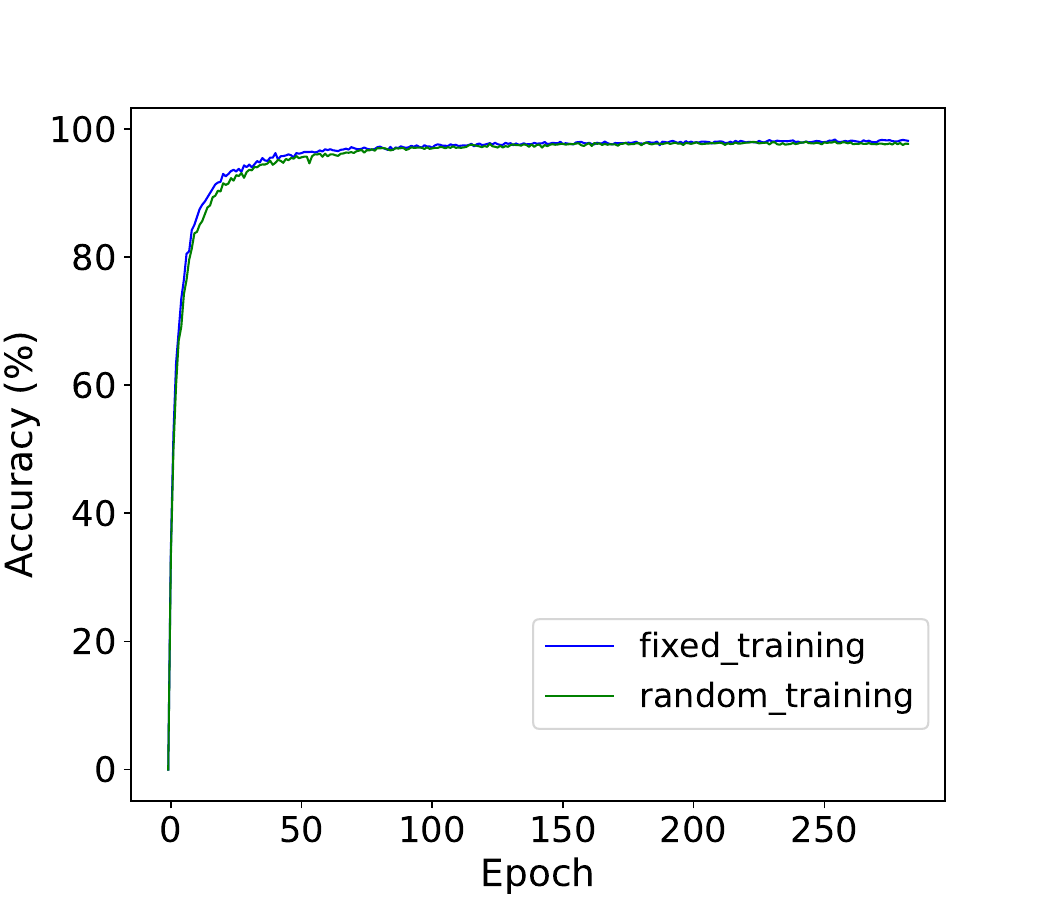}
    \includegraphics[width=0.48\linewidth]{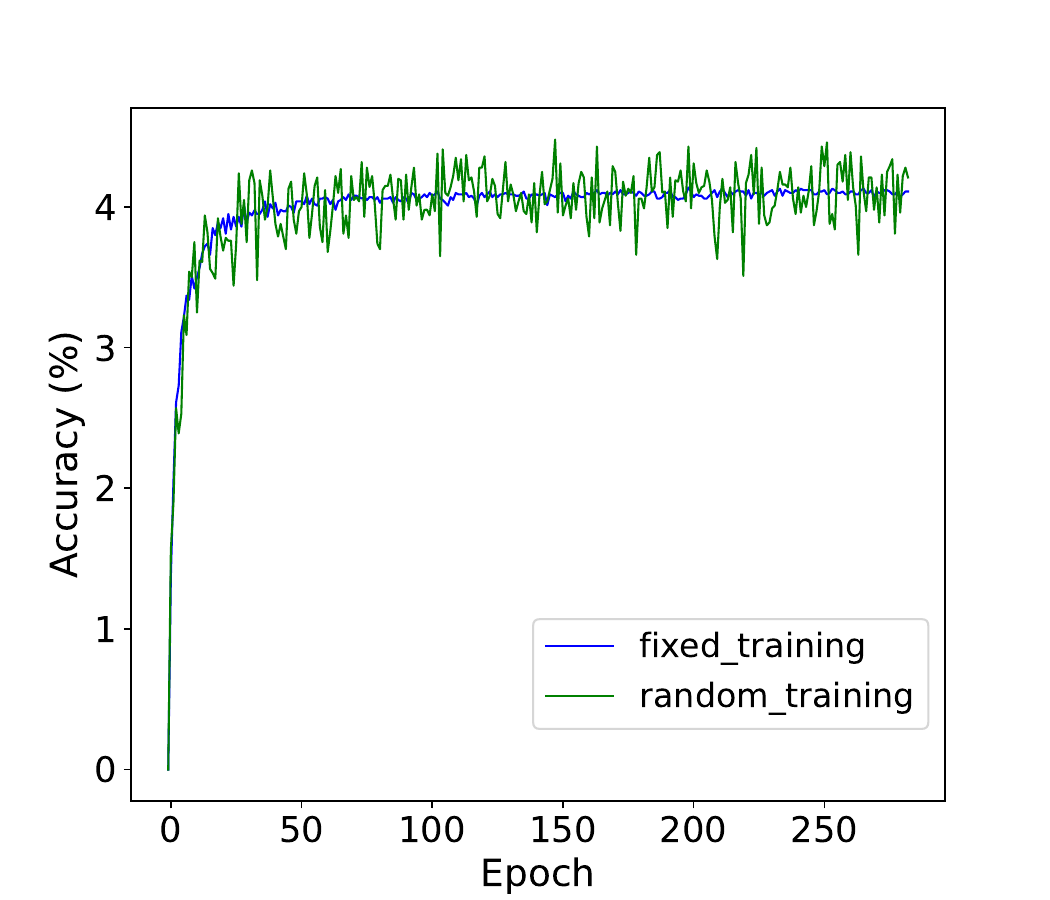}
    \includegraphics[width=0.48\linewidth]{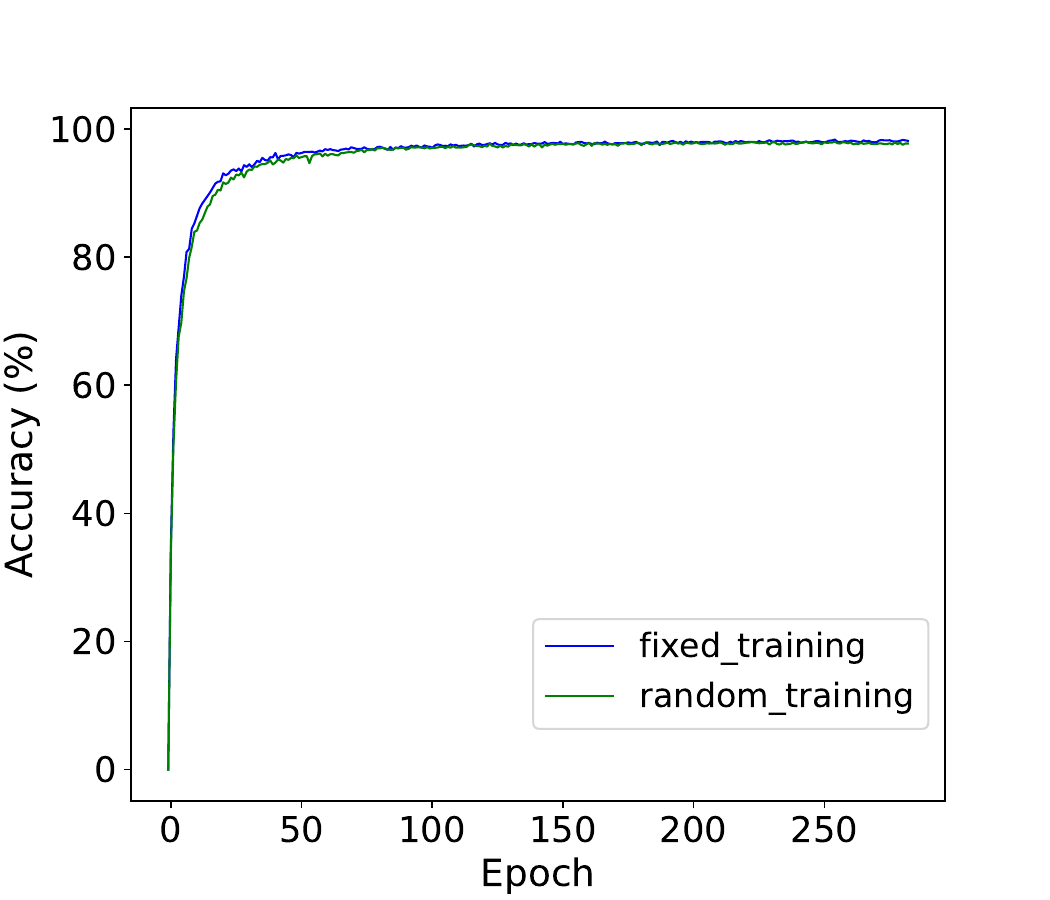}
    \includegraphics[width=0.48\linewidth]{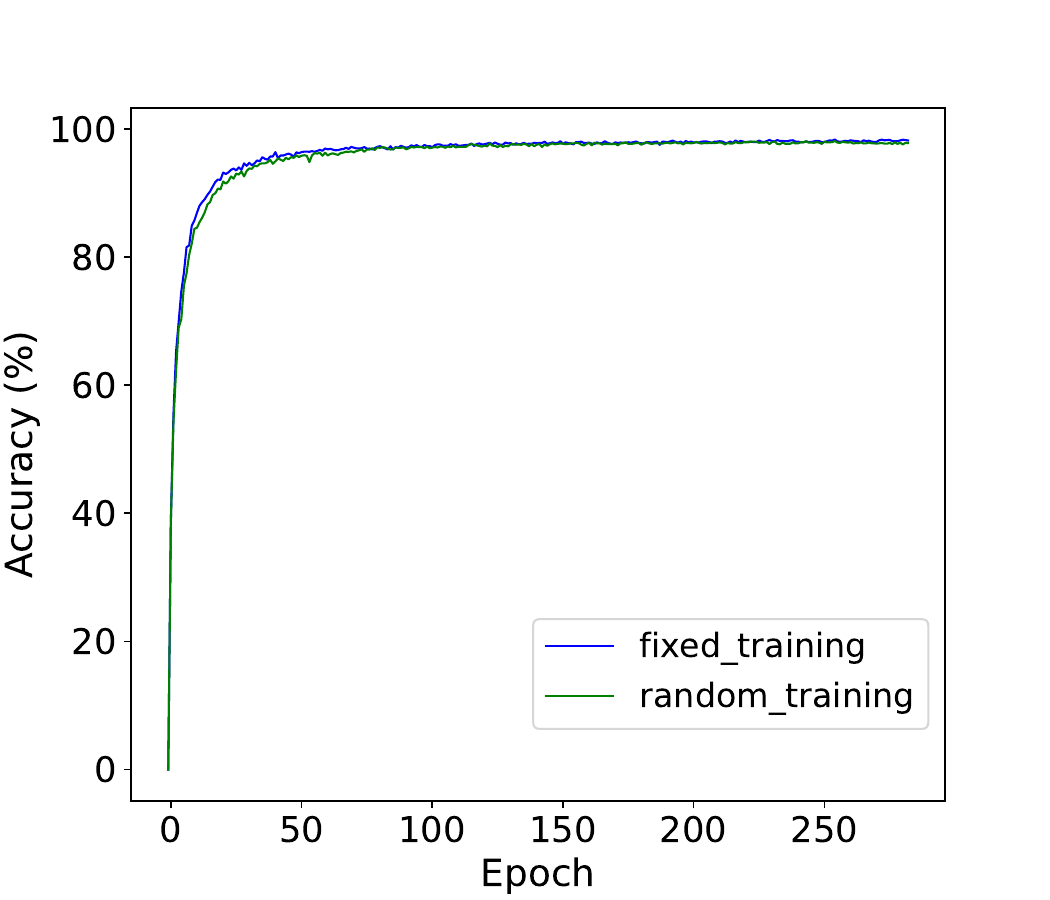}
    \caption{Accuracy comparison between training data with canonical ordered output and training data with randomly permuted output. Top left: permutation-invariant accuracy. Top right: model accuracy. Lower left: central charge accuracy. Lower right: algebra label accuracy.}
    \label{fig:mixVSrand}
\end{figure}

\begin{figure}
    \centering
    \includegraphics[width=.48\textwidth]{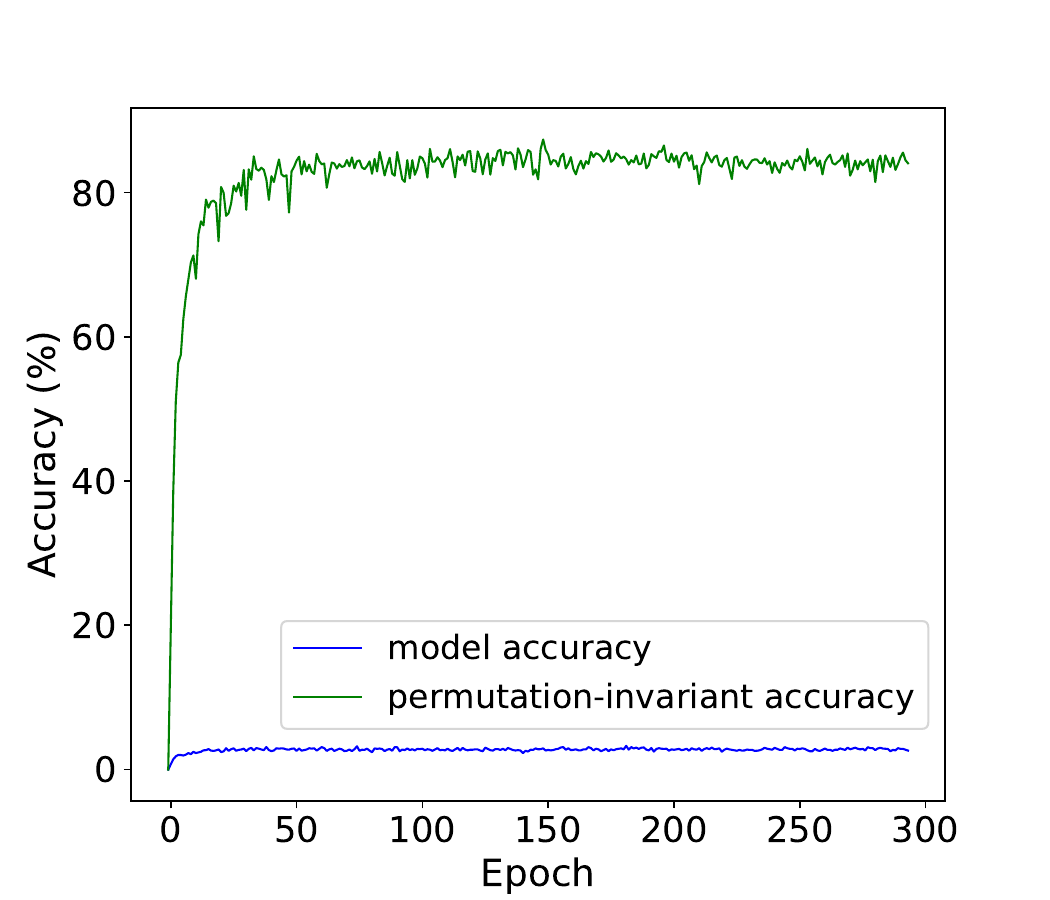}
    \includegraphics[width=.48\textwidth]{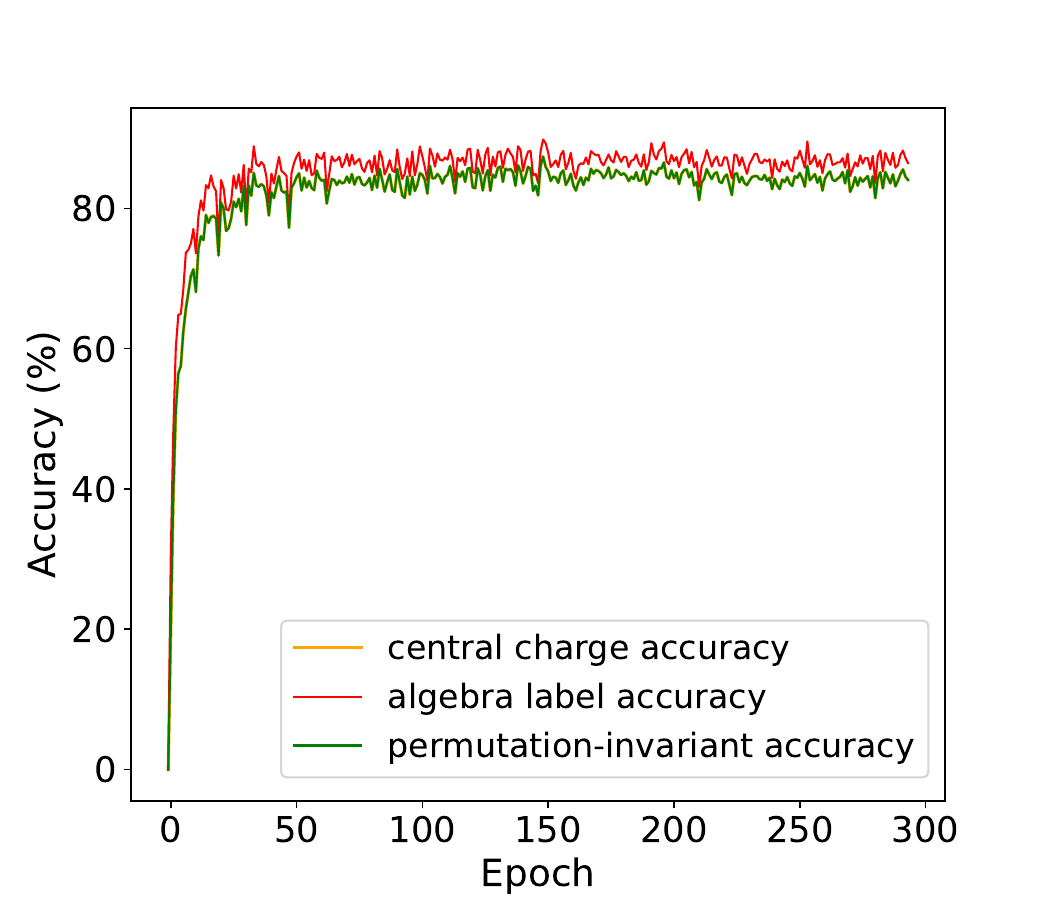}
    \caption{Accuracy curves for tensor product theories with $50 \leq c \leq 100$ trained on tensor product theories with $c<50$. We compare the permutation-invariant accuracy with model accuracy (left). We also compare the permutation-invariant accuracy with the central charge and algebra label accuracy (right).}
    \label{fig:cft_KM_c50-100_acc}
\end{figure}

In this appendix, we include the accuracy curves as supplementary figures for Table \ref{tab: acc_c<50} and Table \ref{tab: acc_c50-100}. In Fig \ref{fig:cft_KM_c50_base_acc}, we monitor the learning dynamics of transformers trained on $3,000,000$ tensor product theories with $c<50$ and tested on a held-out dataset with $10,000$ examples. We have the same conclusion as we have drawn from Table \ref{tab: acc_c<50}: all metrics except model accuracy overlap throughout $300$ epochs. In addition, we plotted the accuracy comparison between training datasets where the outputs are randomly permuted and those where the outputs have a canonical ordering. For both cases, the output of the test datasets are randomly permuted. We see that in Fig \ref{fig:mixVSrand}, the model performance with these two different train datasets are the same. 

In Fig \ref{fig:cft_KM_c50-100_acc}, we monitor the learning dynamics of transformers trained on $3,000,000$ tensor product theories with $c<50$ and tested on $10,000$ tensor product theories with $50 \leq c \leq 100$. We also have the same conclusion as in Table \ref{tab: acc_c50-100}: algebra label accuracy converges a little faster than rest of the metrics and reaches a higher plateau.

\end{document}